\documentclass[iop,emulateapj-rtx4]{emulateapj}
\usepackage{apjfonts}
\usepackage{amssymb,latexsym,amsmath,graphics}
\slugcomment{} \shorttitle{ON THE HIGH-FREQUENCY QPOs FROM BLACK
HOLES} \shortauthors{ERKUT}

\begin{document}

\title{On the high-frequency quasi-periodic oscillations from black holes}

\author{M. Hakan Erkut\altaffilmark{}}

\affil{\altaffilmark{}Department of Physics, \.Istanbul
K\"{u}lt\"{u}r University, Atak\"{o}y Campus, Bak\i rk\"{o}y 34156,
\.Istanbul, Turkey}

\altaffiltext{}{m.erkut@iku.edu.tr}

\begin{abstract}
We apply the global mode analysis, which has been recently developed
for the modeling of kHz quasi-periodic oscillations (QPOs) from
neutron stars, to the inner region of an accretion disk around a
rotating black hole. Within a pseudo-Newtonian approach that keeps
the ratio of the radial epicyclic frequency $\kappa $ to the orbital
frequency $\Omega$ the same as the corresponding ratio for a Kerr
black hole we determine the innermost disk region where the
hydrodynamic modes grow in amplitude. We find that the radiation
flux emerging from the inner disk has the highest values within the
same region. Using the flux weighted averages of the frequency bands
over this region we identify the growing modes with highest
frequency branches $\Omega +\kappa $ and $\Omega $ to be the
plausible candidates for the high-frequency QPO pairs observed in
black hole systems. The observed frequency ratio around $1.5$ can
therefore be understood naturally in terms of the global free
oscillations in the innermost region of a viscous accretion disk
around a black hole without invoking a particular resonance to
produce black hole QPOs. Although the frequency ratio $\left\langle
\Omega +\kappa \right\rangle $/$\left\langle \Omega \right\rangle $
is found to be not sensitive to the black hole's spin which is good
for explaining the high-frequency QPOs it may work as a limited
diagnostic of the spin parameter to distinguish black holes with
very large spin from the slowly rotating ones. Within our model we
estimate the frequency ratio of a high-frequency QPO pair to be
greater than $1.5$ if the black hole is a slow rotator. For fast
rotating black holes, we expect the same ratio to be less than
$1.5$.
\end{abstract}


\section{Introduction\label{intr}}

The high energy emission from neutron stars and black holes in X-ray
binaries is generally powered by accretion onto the compact object.
The variability of X-ray light curve with different time scales from
milliseconds to days is usually attributed to various characteristic
time scales associated with accretion flow around the black hole or
neutron star. In low-mass X-ray binaries (LMXBs), where the central
object is fed by an accretion disk, a few variability frequencies
observed as quasi-periodic oscillation (QPO) peaks in addition to
other broad-band features in the power spectra are common to both
black hole and neutron star sources. Although there are some
phenomenological differences between QPOs in black hole candidates
and those observed in neutron star LMXBs, the similarities such as
tight correlations of high and low frequency power spectral features
in black hole and neutron star sources are remarkable (van der Klis
1994; Psaltis, Belloni, \& van der Klis 1999; Wijnands \& van der
Klis 1999). Any interpretation or model solely based on the
existence of a magnetic field, a hard surface, which are neutron
star like properties, or an innermost stable circular orbit (ISCO)
as a black hole like property to produce QPOs cannot account for the
correlations of timing properties among different sources.

QPOs were discovered in the X-ray power density spectrum of black
hole transients with frequencies in the $\simeq 0.1-30$ Hz range
during the very high spectral state of these sources (Motch et al.
1983; Miyamoto \& Kitamoto 1989; Miyamoto et al. 1991; Morgan,
Remillard, \& Greiner 1997; Wijnands, Homan, \& van der Klis 1999;
Sobczak et al. 2000; Strohmayer 2001; Muno et al. 2001). For several
black hole sources, high-frequency QPOs were observed as single
peaks roughly around $200$ Hz (see Remillard \& McClintock 2006 and
references therein). The discovery of twin kHz QPOs in neutron star
LMXBs with their peak frequencies within around the $200-1200$ Hz
range and a peak separation in the $\simeq 180-360$ Hz range
together with their tight correlations with the low frequency power
spectral components which are also observed in black hole candidates
strengthens the idea that QPOs are produced in the inner regions of
accretion disks around compact objects (see van der Klis 2000 and
references therein; M\'{e}ndez \& Belloni 2007). The discoveries of
twin hectoHz QPOs from black hole candidates such as GRO J1655$-$40,
XTE J1550$-$564, and GRS 1915$+$105 (Remillard et al. 2002, 2003;
Remillard \& McClintock 2006) have almost certified the idea of
unifying the interpretation of high-frequency QPOs observed in black
hole and neutron star sources within a single QPO model.

One of the most striking differences between the high-frequency QPOs
of black holes and the kHz QPOs in neutron star LMXBs is the fact
that the former do not show any significant correlation with the
X-ray luminosity whereas the latter do. From one observation to
another in a given source, the frequency shifts in the
high-frequency QPOs of black holes are negligible as compared to the
variations in the frequencies of kHz QPO peaks of accreting neutron
stars. As compared to kHz QPOs detected in the power spectra of
neutron stars, high-frequency QPOs from black holes are weak
features having relatively low-quality factors. The commonly
observed property of both low and high-frequency QPOs in black holes
is the fact that these oscillations are strongest at photon energies
above 6 keV when the power law component in the energy spectra
dominates over the disk component (Remillard \& McClintock 2006).
Though it is not conclusive regarding the number of black hole
sources exhibiting twin QPOs, the frequency ratio of the upper
high-frequency QPO to the lower one is close to 3:2 in black hole
candidates. There is no specific value for the ratio of two
simultaneous kHz QPOs observed in neutron star sources; the ratio of
the upper QPO frequency to the lower one rather takes different
values changing between one and three from one source to another
(Belloni, M\'{e}ndez, \& Homan 2005). Beside similarities and tight
correlations, the phenomenological differences between the
high-frequency QPOs observed from black hole candidates and those
from neutron star LMXBs likely arise from the dominant effect of
different boundary conditions imposed by a hard surface or a
magnetic field and the ISCO as they might be more appropriate for a
neutron star and a black hole, respectively.

High-frequency QPOs were detected in seven black hole sources among
which three black hole candidates exhibited QPO pairs in their power
spectra. In all the sources with QPO pairs of commensurate
frequencies, the ratio of two QPO frequencies is very close to 1.5
in two black hole binaries and to 1.6 in the third one (Remillard et
al. 2002; Remillard et al. 2003). The common property of
high-frequency QPOs observed in black hole systems is based on the
spectral state of a given source. All high-frequency black-hole QPOs
are usually observed in very high and high spectral states which are
characterized by the most luminous states of the source (see
Remillard \& McClintock 2006 and references therein). For such high
luminosities, the accretion disk around the black hole is expected
to be truncated at the radius of the ISCO according to the standard
model (Novikov \& Thorne 1973; Shakura \& Sunyaev 1973). For radii
less than the radius of the ISCO, the accreting gas is thought to
plunge radially towards the black hole. In the recent MHD
simulations by Beckwith, Hawley, \& Krolik (2008), however, the
innermost ring of the disk which emits significant radiation has
been shown to lie inside where the standard model predicts. These
simulations have modified the stress-free boundary condition of the
standard model at the ISCO, but they have employed the classical
relationship between magnetic stresses and energy dissipation and
the assumption that the radial inflow time-scale of the accreting
matter inside the ISCO is longer than the time-scales for
thermalization and radiation of the dissipated heat, both of which
may not be valid at all within the plunging region. We therefore
anticipate the innermost part of the disk beyond the ISCO to be the
most probable region near the black hole for the production of
high-frequency QPOs. It is very likely within this region that the
frequencies of the unstable growing disk modes correspond to the
frequency bands of QPOs.

The early attempts to interpret QPO frequencies in terms of disk
modes were made by Alpar et al. (1992) and Alpar \& Y\i lmaz (1997).
The initial contributions to the theory of disk oscillations
explored the role of trapped disk oscillations in the variability of
the X-ray power spectra of black hole sources (Kato \& Fukue 1980;
Nowak \& Wagoner 1991; Kato 2001 and references therein). The
general relativistic test-particle frequencies were recognized and
employed to construct models of QPOs for both neutron stars and
black holes (Stella, Vietri, \& Morsink 1999; Abramowicz et al.
2003). The model of Psaltis \& Norman (2000) revealed the importance
of hydrodynamic corrections to relativistic test-particle
frequencies and the effect of hydrodynamic disk parameters on the
correlations of QPOs and broad-band noise component. Alpar \&
Psaltis (2008) noted that the radial epicyclic frequency would be
the highest dynamical frequency in the inner region of an accretion
disk. This conclusion is based upon the existence of a
magnetohydrodynamic boundary region around the neutron star where
orbital frequencies deviate from Keplerian test-particle frequencies
due to viscous and magnetic stresses (see Erkut \& Alpar 2004). The
recent analysis by Erkut, Psaltis, \& Alpar (2008) of global
hydrodynamic modes in the boundary regions of neutron stars showed
how important the hydrodynamic effects are in estimating the
observational characteristics of high-frequency QPOs.

In this paper we apply the mode analysis, which has been developed by Erkut,
Psaltis, \& Alpar (2008) for the boundary region model of kHz QPOs from
neutron stars, to the inner region of an accretion disk around a Kerr black
hole. In order to account for the general relativistic effects of a Kerr
metric on the ratio of dynamical frequencies, we work with a new
pseudo-Newtonian potential that is appropriate for the analysis of temporal
behavior of relativistic disks. This approach allows us to extend our recent
study of global hydrodynamic modes of free oscillations to black holes as
well. Our aim is to identify the modes whose frequency bands correspond to
the high-frequency QPO pairs usually detected with a frequency ratio close
to 1.5 in black hole sources. Most importantly, we provide a way to use the
frequency ratio of these modes as a diagnostics of the spin parameter of a
rotating black hole that exhibit high-frequency QPOs.

In Section 2 we introduce our pseudo-Newtonian approach. The basic
equations and parameters related to the analysis of disk modes for a
Kerr black hole are presented in Section 3. In Section 4 we come up
with the mode analysis and identify the modes relevant to the
commensurate high-frequency QPO pairs with a frequency ratio around
1.5. We discuss the results and present our conclusions in Section
5.

\section{Pseudo-Newtonian Treatment of Frequencies\label{pnt}}

Using pseudo-Newtonian potentials in hydrodynamic simulations is an
effective and easy method to incorporate relativistic effects into
accretion flows (e.g., Chan, Psaltis, \& \"{O}zel 2009). The
pseudo-Newtonian potential proposed by Paczy\'{n}ski \& Wiita (1980)
is successful in estimating the ISCO for Schwarzschild geometry. It
is therefore appropriate for a non-rotating black hole. Recently,
two modified Newtonian force models have been introduced by
Mukhopadhyay \& Misra (2003) to approximate the dynamical
frequencies of an accretion disk around a rotating black hole.

In our current analysis, the ratio of two successive frequency bands
of disk modes is of interest in identifying the high-frequency QPOs
from black holes. Unlike the early studies we mentioned above our
pseudo-Newtonian treatment of dynamical disk frequencies keeps the
ratio of the radial epicyclic frequency $\kappa $ to the orbital
frequency $\Omega $ exactly the same as the corresponding ratio
observed by a distant observer of a Kerr black hole.

The Kerr expression for the ratio of the test-particle frequencies
$\kappa $ and $\Omega $ is
\begin{equation}
\frac{\kappa }{\Omega }=\sqrt{1-6\left( \frac{R_{\mathrm{g}}}{r}\right)
+8a\left( \frac{R_{\mathrm{g}}}{r}\right) ^{3/2}-3a^{2}\left( \frac{R_{%
\mathrm{g}}}{r}\right) ^{2}},  \label{kker}
\end{equation}%
where $a$ is the spin parameter of the black hole and $R_{\mathrm{g}%
}=GM/c^{2}$ with $M$ and $c$ being the mass of the black hole and the speed
of light, respectively. The Newtonian expression for the same ratio is%
\begin{equation}
\frac{\kappa }{\Omega }=\sqrt{2\left( 2+\frac{d\ln \Omega }{d\ln r}\right) }.
\label{knew}
\end{equation}%
The ratio given by equation (\ref{knew}) is valid for both the
test-particles and the hydrodynamical fluids rotating in orbits. In
the steady state of a geometrically thin disk, the hydrodynamical
effects of pressure gradients on the test-particle frequencies are
negligible. As we mention in Section 3, the radial momentum balance
in a geometrically thin disk can be approximated by the
test-particle orbits where the centripetal acceleration of each gas
particle rotating with the frequency $\Omega $ is due to the
gravitational force. Setting equations (\ref{kker}) and (\ref{knew})
equal to each other, we obtain within our pseudo-Newtonian approach
a differential equation for the orbital frequency $\Omega (r)$.
Given a suitable pseudo-Newtonian potential, the orbital frequency
$\Omega $ satisfies the radial momentum equation and mimics the
effect of strong gravity on the gas-particle orbits by keeping the
value of $\kappa /\Omega $ the same as the corresponding Kerr value
in the test-particle regime. In this sense, our approach is similar
to the early pseudo-Newtonian treatments of the dynamical
frequencies in a geometrically thin accretion disk around a black
hole. To find a solution for $\Omega (r)$, we require that our
pseudo-Newtonian orbital frequency match the Kerr orbital frequency
in the outer disk. According to a distant observer, the Kerr
expression for the orbital frequency is%
\begin{equation}
\Omega _{\mathrm{Kerr}}(r)=\frac{\Omega _{\mathrm{K}}(r)}{1+a\left( R_{%
\mathrm{g}}/r\right) ^{3/2}},  \label{oker}
\end{equation}%
where $\Omega _{\mathrm{K}}(r)=\left( GM/r^{3}\right) ^{1/2}$ is the
Keplerian frequency. For sufficiently large radii, that is for $r\gg R_{%
\mathrm{g}}$, the Kerr orbital frequency assumes its Keplerian value. Using
the same asymptotic boundary condition on the pseudo-Newtonian orbital
frequency, it follows from equations (\ref{kker}) and (\ref{knew}) that%
\begin{equation}
\Omega (r)=\Omega _{\mathrm{K}}(r)\exp \left[ 3\left( \frac{R_{\mathrm{g}}}{r%
}\right) -\frac{8}{3}a\left( \frac{R_{\mathrm{g}}}{r}\right) ^{3/2}+\frac{3}{%
4}a^{2}\left( \frac{R_{\mathrm{g}}}{r}\right) ^{2}\right] .  \label{pnom}
\end{equation}

To illustrate our treatment of the radial epicyclic and orbital
frequencies, we plot in Figure~$1$ the pseudo-Newtonian frequencies
$\kappa $ and $\Omega $ over a wide range of disk radii in
comparison with the corresponding Kerr frequencies. Figure~$1a$
shows the radial profiles of $\kappa $ and $\Omega $ for a
Schwarzschild black hole for which the spin parameter $a=0$. The
radial profiles of the same frequencies for a Kerr black hole with a
spin parameter $a=0.8$ are shown in Figure~$1b$. The
pseudo-Newtonian frequencies $\kappa $ and $\Omega $ can be seen to
slightly deviate from the Kerr frequencies in the inner disk while
they asymptotically match them in the outer disk. Note, however,
that the pseudo-Newtonian frequency $\kappa $ always yields the
correct estimation for the ISCO as $\kappa /\Omega $ matches its
Kerr value exactly for all disk radii.

In the next section, we present the basic equations for a geometrically thin
disk and obtain within the current pseudo-Newtonian approach the
hydrodynamic parameters that are necessary for the analysis of global modes
in the inner disk.

\section{Basic Equations and Parameters\label{ep}}

The long-wavelength global hydrodynamic modes of free oscillations have been
recently studied for the boundary regions of accretion disks around neutron
stars (see Erkut, Psaltis, \& Alpar 2008, hereafter EPA08). In the mode
analysis by EPA08 the basic disk equations are perturbed for a geometrically
thin disk in vertical hydrostatic equilibrium. The Fourier decomposition of
the linearized perturbation equations leads to the identification of the
complex mode frequencies $\omega ^{(m)}$. The real parts of axisymmetric $%
(m=0)$ and nonaxisymmetric $(m\geq 1)$ mode frequencies correspond to the
frequency bands of QPOs while the imaginary parts determine the growth rates
of the oscillations.

In the global mode analysis, both the frequency bands and their growth rates
depend on several key parameters such as $\kappa /\Omega $, the radial
profile of the surface density, $\beta $, the inverse timescale $\Omega
_{\nu }$ associated with the radial drift velocity, and the inverse
timescale $\Omega _{s}$ associated with the sound speed in the inner disk.
These parameters are determined by the global structure of the unperturbed
steady disk (see EPA08).

For sufficiently high mass accretion rates, e.g., $\dot{M}\gtrsim 0.1\dot{M}%
_{\mathrm{E}}$, where $\dot{M}_{\mathrm{E}}$ is the Eddington mass accretion
rate, black holes in LMXBs accrete matter through radiatively efficient
accretion disks whose innermost regions are dominated by radiation pressure
(Shakura \& Sunyaev 1973, hereafter SS73). The innermost truncation radius
of such a disk around a Kerr black hole is estimated by the radius of the
ISCO, $r_{\mathrm{in}}$, which can be found as a solution of $\kappa /\Omega
=0$ for $r=r_{\mathrm{in}}$ (see eq. [\ref{kker}]). The unperturbed steady
structure of a radiation pressure dominated inner disk is described by%
\begin{equation}
\rho c_{s}^{2}=\frac{\varepsilon }{3},  \label{rpd}
\end{equation}%
where $c_{s}$ is the effective sound speed, $\rho $ is the average mass
density, and $\varepsilon $ is the radiation energy density. We write, for
the vertical hydrostatic equilibrium in the disk,%
\begin{equation}
c_{s}^{2}=\frac{1}{2}\Omega ^{2}H^{2},  \label{ss}
\end{equation}%
where $H$ is the half-thickness of the disk. The average mass density can be
expressed in terms of the surface mass density $\Sigma $ as%
\begin{equation}
\rho =\frac{\Sigma }{2H}.  \label{md}
\end{equation}%
The vertical energy balance in the inner disk is satisfied for%
\begin{equation}
\varepsilon =\left( \frac{3\kappa _{\mathrm{es}}\Sigma }{4c}\right) \Phi ,
\label{red}
\end{equation}%
where $\kappa _{\mathrm{es}}$ is the electron scattering opacity and $\Phi $
is the energy dissipation rate per unit area of the disk (see SS73). The
energy flux due to viscous energy dissipation is%
\begin{equation}
\Phi =\frac{1}{2}\nu \Sigma \left( r\frac{d\Omega }{dr}\right) ^{2}.
\label{ef}
\end{equation}%
Here, $\nu $ is the kinematic viscosity for which the $\alpha $-prescription
(SS73) can be written as%
\begin{equation}
-\alpha \Sigma c_{s}^{2}=\nu \Sigma r\frac{d\Omega }{dr}.  \label{ap}
\end{equation}

For a geometrically thin disk, the radial momentum balance can be
written to a good approximation as $\Omega ^{2}r-d\Gamma /dr=0$,
where $\Gamma $ is the pseudo-Newtonian potential that mimics the
gravitational field of the black hole. In the radial momentum
equation, the pseudo-Newtonian force, $-d\Gamma /dr$, is the source
of acceleration, $-\Omega ^{2}r$, where $\Omega $ is given by
equation (\ref{pnom}). For the conservation of mass and angular
momentum, we write
\begin{equation}
-2\pi r\Sigma v_{r}=\dot{M}  \label{cnt}
\end{equation}%
and
\begin{equation}
2\pi \nu \Sigma r^{3}\frac{d\Omega }{dr}+\dot{M}r^{2}\Omega =C,  \label{agm}
\end{equation}%
respectively, where $v_{r}$ is the radial drift velocity of the accreting
matter in the inner disk and $C$ is an arbitrary constant of integration.

We solve equation (\ref{agm}) using torque-free boundary condition
at the innermost disk radius, $r_{\mathrm{in}}$, which is
appropriate for a disk around a black hole. The constant of
integration can be determined as $C=\dot{M}r_{\mathrm{in}}^{2}\Omega
(r_{\mathrm{in}})$ to satisfy the torque-free boundary condition.
Using equation (\ref{pnom}), it follows from equations
(\ref{rpd})--(\ref{agm}) that
\begin{equation}
\beta \equiv \frac{d\ln \Sigma }{d\ln r}=\frac{3}{2}-B(r)-\frac{D(r)}{A(r)}-%
\frac{E(r)}{F(r)},  \label{beta}
\end{equation}%
\begin{equation}
\frac{\Omega _{\nu }}{\Omega _{s}}\equiv -\frac{v_{r}/r}{c_{s}/r}\simeq
2.57\alpha \dot{m}\left( \frac{6R_{\mathrm{g}}}{r}\right) A(r),  \label{nuos}
\end{equation}%
\begin{equation}
\frac{\Omega _{s}}{\Omega }\simeq 2.4\dot{m}\left( \frac{6R_{\mathrm{g}}}{r}%
\right) A(r)F(r),  \label{soo}
\end{equation}%
and%
\begin{equation}
\frac{\Phi }{\Phi _{\mathrm{t}}}=\left( \frac{r}{r_{\mathrm{in}}}\right)
^{-3}f^{2}(r)A(r)F(r),  \label{flx}
\end{equation}%
where $\Phi _{\mathrm{t}}\equiv 3GM\dot{M}/8\pi r_{\mathrm{in}}^{3}$
is the typical value for the radiation flux and
\begin{equation}
\dot{m}\equiv
\frac{\dot{M}}{\dot{M}_{\mathrm{E}}}=\frac{\dot{M}}{1.9\times
10^{18}\mathrm{g}\,\mathrm{s}^{-1}}\left( \frac{\eta }{0.06}\right)
\left( \frac{M}{M_{\odot }}\right) ^{-1}. \label{mdt}
\end{equation}
Here, $\eta $ is a function of the spin parameter $a$ such that
\begin{equation}
a=\frac{4\sqrt{2}\left[ 1-\left( 1-\eta \right) ^{2}\right]
^{1/2}-2\left( 1-\eta \right) }{3\sqrt{3}\left[ 1-\left( 1-\eta
\right) ^{2}\right] } \label{earlt}
\end{equation}
for prograde accretion disks around rotating black holes (see
Shapiro \& Teukolsky 1983). In equations (\ref{beta})--(\ref{flx}),
the dimensionless factors arising from the boundary conditions and
pseudo-Newtonian corrections are
\begin{equation}
A(r)=1+2\left( \frac{R_{\mathrm{g}}}{r}\right) -\frac{8}{3}a\left( \frac{R_{%
\mathrm{g}}}{r}\right) ^{3/2}+a^{2}\left( \frac{R_{\mathrm{g}}}{r}\right)
^{2},  \label{a}
\end{equation}%
\begin{equation}
B(r)=-3\left( \frac{R_{\mathrm{g}}}{r}\right) +4a\left( \frac{R_{\mathrm{g}}%
}{r}\right) ^{3/2}-\frac{3}{2}a^{2}\left( \frac{R_{\mathrm{g}}}{r}\right)
^{2},  \label{b}
\end{equation}%
\begin{equation}
D(r)=-4\left( \frac{R_{\mathrm{g}}}{r}\right) +8a\left( \frac{R_{\mathrm{g}}%
}{r}\right) ^{3/2}-4a^{2}\left( \frac{R_{\mathrm{g}}}{r}\right) ^{2},
\label{d}
\end{equation}%
\begin{equation}
E(r)=\frac{f(r_{\mathrm{in}})}{f(r)}\left( B(r)+\frac{1}{2}\right) \left(
\frac{r}{r_{\mathrm{in}}}\right) ^{-1/2},  \label{e}
\end{equation}%
and%
\begin{equation}
F(r)=1-\frac{f(r_{\mathrm{in}})}{f(r)}\left( \frac{r}{r_{\mathrm{in}}}%
\right) ^{-1/2},  \label{cf}
\end{equation}%
with%
\begin{equation}
f(r)=\exp \left[ 3\left( \frac{R_{\mathrm{g}}}{r}\right) -\frac{8}{3}a\left(
\frac{R_{\mathrm{g}}}{r}\right) ^{3/2}+\frac{3}{4}a^{2}\left( \frac{R_{%
\mathrm{g}}}{r}\right) ^{2}\right] .  \label{f}
\end{equation}

For illustrative purposes we display in Figure~$2$ the radial
distributions of the outgoing radiation flux (see eq. [\ref{flx}])
throughout the inner disk for two putative black holes with spin
parameters $a=0$ and $a=0.9$. In
the following section, we use the ratio $\kappa /\Omega $ (see eq. [\ref%
{kker}]) and the global hydrodynamic parameters $\beta $, $\Omega
_{\nu}/\Omega _{s}$, and $\Omega _{s}/\Omega $ (see eq. [\ref{beta}]--[\ref%
{soo}]) to identify the radial zone in the inner disk where the
modes grow. As we will see, the hydrodynamic modes grow only within
a limited range of radii in the innermost disk region out of which
the radiation flux is maximum (see Fig.~$2$).

\section{Global Modes in the Inner Disk\label{mode}}

When there are no external perturbations due to large-scale magnetic fields
of the accreting star, the free oscillation modes in a boundary region or
the inner disk are excited through the dynamical effect of the viscosity. In
the limit of small hydrodynamic corrections, this can be seen from the
growth rates of both axisymmetric and nonaxisymmetric modes for which Im$%
\left( \omega ^{(m)}\right) \propto -\beta \Omega _{\nu }$ (see
EPA08). In the presence of viscosity and therefore of radial drift
velocity, $\Omega _{\nu }\neq 0$ and the high-frequency modes can
have positive growth rates only if $\beta <0$. This is also valid
for the global modes in a disk around a black hole. Unlike neutron
stars, the effect of a large-scale toroidal magnetic force in
addition to viscosity on the excitation of global modes might be
absent in the case of black holes (see Section 3 in EPA08).

For a black hole disk, the presence of the ISCO with a torque-free boundary
condition determines the radial profile of the surface density, $\beta $,
and thus the growth rates of the modes. Note that $\beta \simeq 3/2>0$ for $%
r\gg r_{\mathrm{in}}$ (see eq. [\ref{beta}]) and we expect, in the regime of
small hydrodynamic corrections, that the global hydrodynamic modes do not
grow for sufficiently large radii in the inner disk. For the innermost disk
region, however, the hydrodynamic corrections can be important to
distinguish among the growth rates of different modes and to identify the
set of radii at which these modes grow. In order to see the effects of
hydrodynamic parameters on both the frequency bands and growth rates of the
modes in the inner disk beyond the regime of negligible hydrodynamic
corrections, we use equations (\ref{beta})--(\ref{soo}) together with
equation (\ref{kker}) in the full eigenfrequency solutions for axisymmetric
and nonaxisymmetric perturbations given in the Appendix of EPA08.

Figures~$3$--$6$ show the real and imaginary parts of the complex
mode frequencies in units of the orbital frequency $\Omega $ as
functions of the radial distance in the inner disk. The real and
imaginary parts represent the frequency bands and the growth rates
of the modes, respectively. In Figures~$3$--$6$, we label the
hydrodynamic mode frequencies and their growth rates with notation
corresponding to the test-particle frequencies. This provides us
with an easy identification and a simple designation of each mode
without ambiguity and without loss of generality. In the limit of
small hydrodynamic corrections, the frequencies of all hydrodynamic
modes converge to the test-particle frequencies. We mark
axisymmetric $\left( m=0\right) $ modes with the corresponding
test-particle frequencies $\omega =0$ and $\omega =\kappa $. We use
$\omega =\Omega $ and $\omega =\Omega \pm \kappa $ as the
appropriate labels to distinguish among nonaxisymmetric modes
$\left( m=1\right) $.

Figure~$3a$ exhibits the run of the mode frequencies in the inner
disk of a Schwarzschild black hole $\left( a=0\right) $. We display
the growth rates
of the modes in Figure~$3b$. Figure~$3$ is obtained for $\alpha =0.1$ and $%
\dot{m}=0.1$. Note that the hydrodynamic modes do not grow in the inner disk
for $r\gtrsim 2.3r_{\mathrm{in}}$. The radial zone within which all modes
grow covers only a limited range of radii around $r_{\mathrm{in}}-2.1r_{%
\mathrm{in}}$ in the innermost disk region, as shown in Figure~$3b$.
The hydrodynamic modes with frequency bands around $\Omega $ and $0$
have relatively higher growth rates as compared to those around
$\Omega \pm \kappa $ and $\kappa $ bands. Figure~$4$ reveals how the
mode frequencies and the growth rates are affected by the rotation
of the black hole. For the same values of the viscosity parameter
and the mass accretion rate, that is, for $\alpha =0.1$ and
$\dot{m}=0.1$, we plot the frequency profiles of the modes in
Figure~$4a$ and the corresponding growth rates in Figure~$4b$ for a
rotating black hole with a spin parameter $a=0.9$. In comparison
with Figure~$3b$, the growth rates of the modes are higher in
Figure~$4b$. The range of
radii at which the hydrodynamic modes grow is around $r_{\mathrm{in}}-2.3r_{%
\mathrm{in}}$. All modes decay for $r\gtrsim 2.5r_{\mathrm{in}}$ (see Fig.~$%
4b$). We obtain Figure~$5$ and Figure~$6$ keeping the spin parameter
of the black hole at $a=0.9$, however, changing the viscosity
parameter $\alpha $ and the mass accretion rate $\dot{m}$. For
$\alpha =0.01$ and $\dot{m}=0.1$,
we explore the run of the mode frequencies and the growth rates in Figures~$%
5a$ and $5b$, respectively. As compared to Figure~$4a$, the radial
profiles of the mode frequencies can be seen to be almost unaffected
by a change in
the viscosity parameter $\alpha $ (see Fig.~$5a$). We observe, in Figure~$5b$%
, that the growth rates are lower than those in Figure~$4b$ by a
factor
around $0.1$ which, indeed, is the factor of decrease in $\alpha $. Figure~$%
6 $ is obtained for $\dot{m}=0.6$ while keeping the values of
$\alpha $ and the spin parameter $a$ the same as in Figure~$5$. Note
that both the frequency bands and the growth rates of the modes are
modified to some level at relatively high mass accretion rates. The
greater the mass accretion rate
$\dot{m}$, the higher are the growth rates of hydrodynamic modes (see Fig.~$%
6b$). We note that the frequency bands that are related to $\Omega
\pm \kappa $, $\Omega $, and $\kappa $ branches in the limit of
small hydrodynamic corrections begin to deviate from the
test-particle frequencies for sufficiently large mass accretion
rates $\left( \dot{m}\gtrsim 0.6\right) $ as shown in Figure~$6a$.

The common property of Figures~$3$--$6$ is that all the hydrodynamic
modes grow within a limited region in the innermost part of the disk
with characteristic radii in the $\simeq
r_{\mathrm{in}}-2.5r_{\mathrm{in}}$ range. The radiation flux
emerging from the same region attains the highest
values with maxima at $r\simeq 1.6r_{\mathrm{in}}$ and $r\simeq 1.7r_{%
\mathrm{in}}$ for the black holes with spin parameters $a=0$ and $a=0.9$,
respectively (see Fig.~$2$). It is interesting to deduce from Figures~$3$--$%
6 $ that the frequency ratio of the hydrodynamic modes we associate with $%
\Omega +\kappa $ and $\Omega $ frequency bands is close to $1.5$ at
radii in the $\simeq (1.6-1.7)r_{\mathrm{in}}$ range, where the disk
flux is maximum. As mentioned above, the modes, however, grow
throughout an extended region, with
$r_{\mathrm{in}}<r<2.5r_{\mathrm{in}}$, of radii rather than being
excited at a particular radius. Moreover, the modes with frequencies
around $\Omega -\kappa $ and $\kappa $ bands also grow within the
same region. To distinguish among the pairs of growing modes which
can be regarded as plausible candidates for the high-frequency QPO
pairs from black holes, we consider the mutual ratios of the flux
weighted averages of the frequency bands for different modes. We
define the flux
weighted average of a frequency branch $\omega $ as%
\begin{equation}
\left\langle \omega \right\rangle =\int_{r_{\mathrm{in}}}^{r_{\mathrm{c}%
}}2\pi r\omega (r)\Phi (r)dr\left/ \int_{r_{\mathrm{in}}}^{r_{\mathrm{c}%
}}2\pi r\Phi (r)dr\right. ,  \label{wav}
\end{equation}%
where $r_{\mathrm{c}}$ is the critical radius beyond which the corresponding
mode decays in the inner disk. Using equation (\ref{wav}), we calculate the
ratios of the flux weighted averages of the frequency bands $\Omega +\kappa $%
, $\Omega $, $\Omega -\kappa $, and $\kappa $ for different values of the
black hole spin parameter $a$ between $0$ and $1$. For each model value of $%
a $, we find a critical radius $r_{\mathrm{c}}$ such that all the
hydrodynamic modes grow for $r\leq r_{\mathrm{c}}$. Unlike the
growth rates, the mode frequencies and the width of the radial zone
where the modes grow are sensitive to the spin parameter $a$, but
not sensitive to the viscosity parameter $\alpha $ and the mass
accretion rate $\dot{m}$ (see Figs.~$3$--$6$). In Figure~$7$ we
display the run of $\left\langle \Omega +\kappa \right\rangle
$/$\left\langle \Omega \right\rangle $, $\left\langle \Omega
\right\rangle $/$\left\langle \kappa \right\rangle $, and
$\left\langle
\Omega \right\rangle $/$\left\langle \Omega -\kappa \right\rangle $ for $%
0\leq a\lesssim 1$. Figure~$7$ is obtained for the typical values,
$\alpha =0.1$ and $\dot{m}=0.1 $. The values for $\left\langle
\Omega +\kappa \right\rangle $/$\left\langle \Omega \right\rangle $
are densely clustered around $1.5$ over a wide range of values for
$a$ as shown in Figure~$7$. For slow rotators $\left( a\simeq
0\right) $, $\left\langle \Omega +\kappa \right\rangle
$/$\left\langle \Omega \right\rangle \simeq 1.6$. The same ratio
drops below $1.5$ as $a$ approaches $1$ for rapidly rotating black
holes. The values of $\left\langle \Omega \right\rangle
$/$\left\langle
\kappa \right\rangle $ and $\left\langle \Omega \right\rangle $/$%
\left\langle \Omega -\kappa \right\rangle $, on the other hand, span
a wide range as the spin parameter $a$ varies between $0$ and $1$
(see Fig.~$7$).
Our analysis suggests the hydrodynamic modes with frequency bands around $%
\Omega +\kappa $ and $\Omega $ to be the plausible candidates for
the high-frequency QPO pairs observed in black hole systems. Note
that our model estimation for the frequency ratio of high-frequency
QPO pairs involves the two highest frequency modes with positive
growth rates.

We give examples for surface density perturbations of such global
modes in Figures~$8$--$11$. We display the three dimensional profile
of surface density perturbation $\Sigma _{1}$ in terms of background
surface density $\Sigma _{0}$ in the innermost region
($r_{\mathrm{in}}\leq r\leq 3r_{\mathrm{in}}$) of a disk around a
rotating black hole with spin parameter $a=0.9$ for the typical
values, $\alpha =0.1$ and $\dot{m}=0.1$. The examples for
nonaxisymmetric modes with frequencies $\Omega $ (Fig.~$8$) and
$\Omega
+\kappa $ (Fig.~$9$) show the surface density perturbations at the time $t=2P_{%
\mathrm{in}}$, where $P_{\mathrm{in}}=2\pi /\Omega
(r_{\mathrm{in}})$ is the rotation period at the innermost disk
radius. In Figures~$8$ and $9$, the spiral like shapes of different
iso-level contours plotted on the $xy$-plane reveal the similar
nonaxisymmetric nature of these modes. In the long run, such as for
$t>100P_{\mathrm{in}}$, the surface density perturbations of both
axisymmetric and nonaxisymmetric modes grow in amplitude only for
$r\lesssim 2r_{\mathrm{in}}$ within the same domain. We illustrate
this typical behavior in Figure~$11$ as compared to Figure~$10$ for
the case of axisymmetric
mode with frequency $\kappa $. Note that the perturbations at the time $%
t=15P_{\mathrm{in}}$ (see Fig.~$10$) are comparable in amplitude
over the whole computational domain ($r_{\mathrm{in}}\leq r\leq
3r_{\mathrm{in}}$). The perturbations at the time
$t=200P_{\mathrm{in}}$, however, have large amplitudes only for
$r_{\mathrm{in}}\leq r<2r_{\mathrm{in}}$ whereas their amplitudes
become negligible for $r\gtrsim 2r_{\mathrm{in}}$ (see Fig.~$11$).

In the global three-dimensional magnetohydrodynamic simulations of
black hole accretion disks, the innermost disk region near the ISCO
was found to show QPOs with frequency around the maximum of
epicyclic frequency (Machida \& Matsumoto 2003). In one of the
recent simulations of the three-dimensional magnetohydrodynamic
accretion flows around Schwarzschild black holes (Kato 2004), the
structure of the flow has been changed at
radial distances within the $3.8\,r_{\mathrm{S}}\leq r\leq 6.3\,r_{\mathrm{S%
}}$ range, where $r_{\mathrm{S}}$ is the Schwarzschild radius. Two
pairs of QPOs have been observed to be excited in that region with
frequencies around the Keplerian frequency and the sum of Keplerian
and epicyclic frequencies in the power spectra of these simulations.
Most importantly, the frequency ratio of these QPO features has been
found to be near 1.5. These results are in close agreement with the
result of our mode analysis in the present work.

\section{Discussion and Conclusions\label{conc}}

We have probed the stability of the global modes in the inner region
of a standard accretion disk around a black hole. Our study is the
application of the recently developed analysis of global
hydrodynamic modes (see EPA08) to the identification of the
high-frequency QPO pairs observed in black hole sources. The
presence of the ISCO allows for effects of strong gravity on both
the dynamical frequencies and the global hydrodynamic parameters.
Our pseudo-Newtonian approach takes account of these effects to
determine the frequency bands and the growth rates of the unstable
modes in the inner disk.

The disk is truncated at the radius of the ISCO, $r_{\mathrm{in}}$. The
growth rates of the modes are negative for sufficiently large distances from
the ISCO. We find that the modes grow in amplitude only within a narrow zone
in the innermost disk region. For a non-rotating black hole $\left(
a=0\right) $ the characteristic radii of the zone lie in the $\simeq r_{%
\mathrm{in}}-2.1r_{\mathrm{in}}$ range. The modes grow within the $\simeq r_{%
\mathrm{in}}-2.5r_{\mathrm{in}}$ range for a rotating black hole with spin $%
a=0.9$. Among the growing modes the growth rates of the frequency branches
around $\Omega $ and $0$ are higher as compared to those of the modes with
frequency bands around $\Omega \pm \kappa $ and $\kappa $ (see Figs.~$3$--$6$%
). Due to the effect of enhanced hydrodynamic corrections on the growth
rates, the modes grow faster in an accretion regime with relatively high
rate and viscosity (see EPA08).

The radiation flux due to viscous energy dissipation in the inner disk takes
the highest values within the narrow region where the modes grow (see Fig.~$%
2 $). We deduce from the radial profiles of the mode frequencies
that the frequency ratio of the modes around $\Omega +\kappa $ and
$\Omega $ bands is very close to $1.5$ at the radius where the disk
radiation is maximum. This value was observed for the frequency
ratio of the high-frequency QPO pairs in black hole sources (see
Remillard \& McClintock 2006). Instead of being excited at a
particular radius in the disk, the hydrodynamic modes grow in a
region of finite radial extension. To make an estimation for the
expectation value of a frequency band and therefore for the
frequency ratios of the relevant modes, we calculate the flux
weighted averages of the frequency bands over the innermost disk
region where the modes grow. Scanning the ratios of the expected
mode frequencies for all possible values of the spin parameter $a$
(see Fig.~$7$), we find that only the modes around $\Omega +\kappa $
and $\Omega $ branches have a frequency ratio around $1.5$. This
ratio is slightly higher than $1.5$ if the black hole is a slow rotator $%
\left( 0\leq a\lesssim 0.8\right) $. The same ratio falls below
$1.5$ for fast rotating black holes $\left( 0.9\lesssim a<1\right)$.
The frequency ratios of other modes significantly deviate from $1.5$
over a wide range of values for the spin parameter. Relying on the
observed values for the frequency ratio of the upper high-frequency
QPO to the lower one, we
conclude that the modes with frequency branches $\Omega +\kappa $ and $%
\Omega $ are the most plausible candidates for the high-frequency QPOs from
black holes.

The observations of high-frequency QPOs can be used to determine the
underlying mechanism that produces these oscillations and to measure
the spin parameter of the black hole (Remillard \& McClintock 2006).
Our analysis may provide a way to employ the observed frequency
ratio of a high-frequency QPO pair in a given source to estimate the
spin parameter $a$. In this sense, Figure~$7$ comes out as an
efficient tool for reading the spin parameter $a$ that corresponds
to the value of $\left\langle \Omega +\kappa \right\rangle
$/$\left\langle \Omega \right\rangle $ to be interpreted as the
frequency ratio of a high-frequency QPO pair observed in the X-ray
power spectra of the black hole source.

There are several reasons for expecting to see the fingerprint of
global long wavelength modes in the form of high-frequency QPOs
observed in the X-ray power spectra of black hole sources in LMXBs.
As compared to the neutron stars, there is little chance for
accretion flows around the black holes in LMXBs to be affected by
the dynamical action of a magnetic field of stellar origin. In the
case of a black hole, instead of a direct feedback from the compact
object, except gravity, the fluctuations in the mass transfer rate
from the binary companion introduce perturbations with a broad band
of frequencies including those of the inner disk. The disk modes
which depend on global disk parameters become unstable in the
innermost disk region and thus the disk oscillation frequencies are
selectively amplified without any need for an external mechanism to
force them to attain high amplitudes. Furthermore, the observable
luminosity variation in the X-ray light curve of a source due to
global free oscillation modes of long wavelength, that is, of
sufficiently large lengthscale is expected to be least affected by
the MHD turbulent eddies of short wavelength. According to our
present analysis, the higher the mass accretion rate $\dot{M}$ and
the greater the viscosity parameter $\alpha $, the higher are the
growth rates of the modes. We therefore expect to observe these
modes particularly in the state of high mass accretion rate and high
viscosity. In such a state, the turbulent disk may also interact
with its corona (see Tagger \& Varni\`{e}re 2006). This would lead
to the formation of high-frequency QPOs in a spectral state where
the contribution from the power law component is important. In our
present analysis we identify the relevant disk modes without
deliberating the disk-corona interaction which we plan to consider
in a future work.

\acknowledgments I would like to express my special thanks to M. A.
Alpar who carefully read the manuscript and contributed it through
various suggestions and to D. Psaltis for reading the manuscript and
very useful discussions. I also thank U. Ertan for his valuable
comments. I would like to thank the anonymous referee whose
suggestions lead me to improve this manuscript. I acknowledge
support from T\"{U}B\.{I}TAK (The Scientific and Technical Research
Council of Turkey) for a postdoctoral fellowship and the Marie Curie
FP6 Transfer of Knowledge Project ASTRONS, MKTD-CT-2006-042722.

$\vspace{3cm}$

\begin{figure}[htbp]
\epsscale{1.0} \plottwo{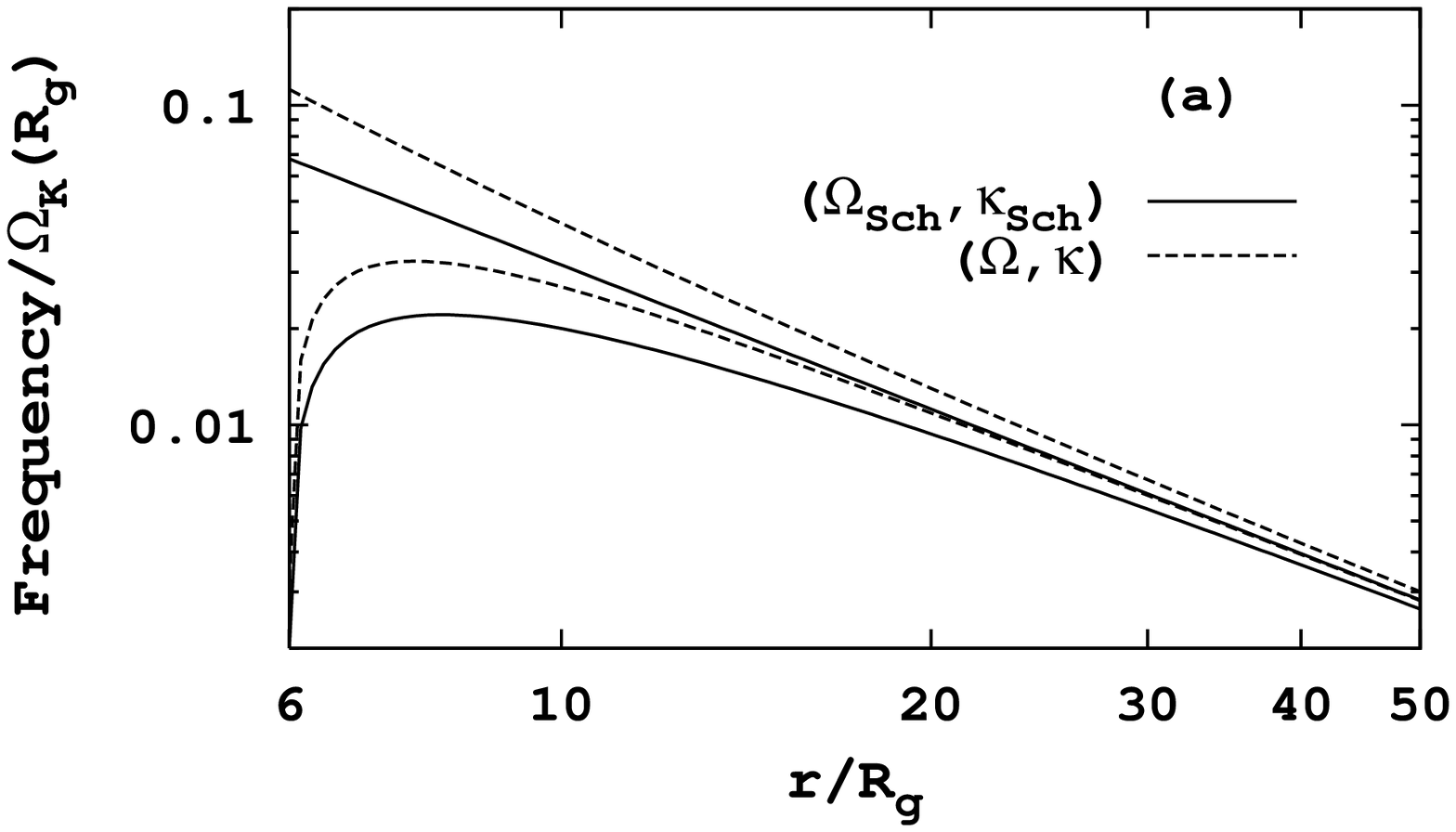}{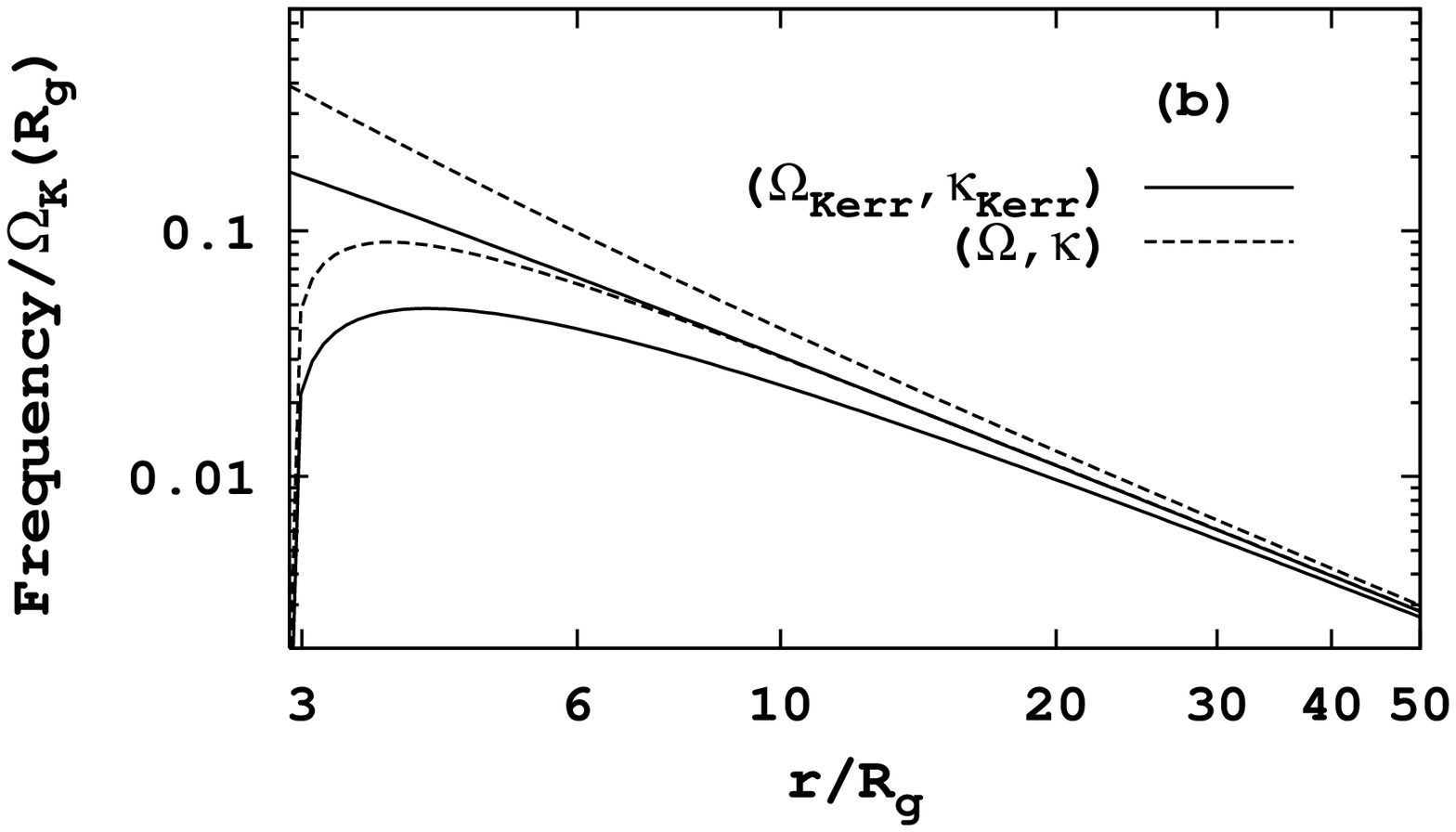} \caption{Radial profiles
of the orbital and epicyclic frequencies. The Schwarzschild orbital
and radial epicyclic frequencies $\Omega _{\mathrm{Sch}}$ and
$\kappa _{\mathrm{Sch}}$ and the Kerr orbital and radial epicyclic
frequencies $\Omega _{\mathrm{Kerr}}$ and $\kappa _{\mathrm{Kerr}}$
are shown by the solid curves. The pseudo-Newtonian orbital
frequency $\Omega $ and the pseudo-Newtonian radial epicyclic
frequency $\kappa $ are shown by the dashed curves. The panel (a) is
obtained for a non-rotating black hole with spin parameter $a=0$.
The frequencies in panel (b) are plotted for a rotating black hole
with spin parameter $a=0.8$. \label{fig1}}
\end{figure}


\begin{figure}[htbp]
\epsscale{1.0} \plottwo{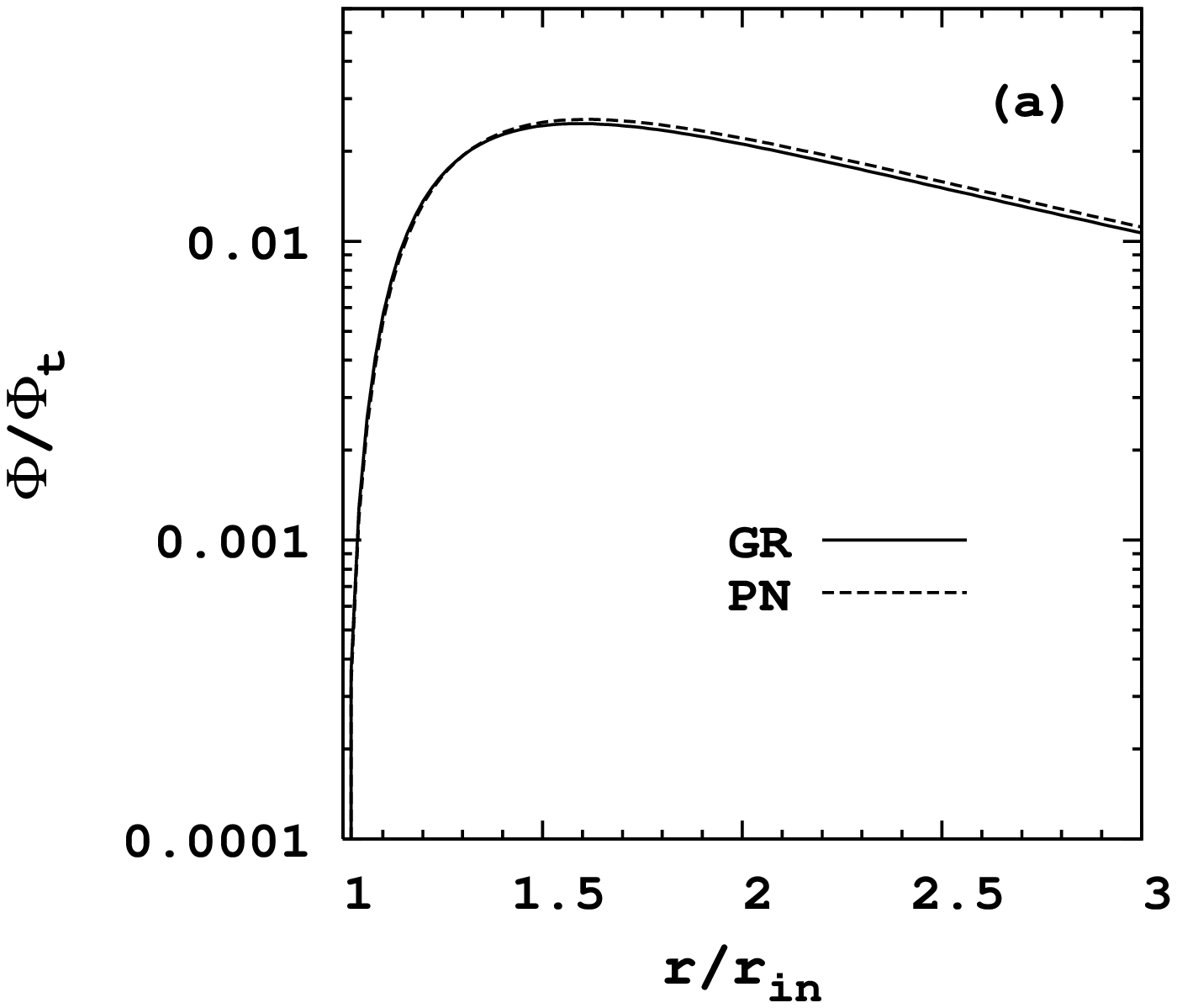}{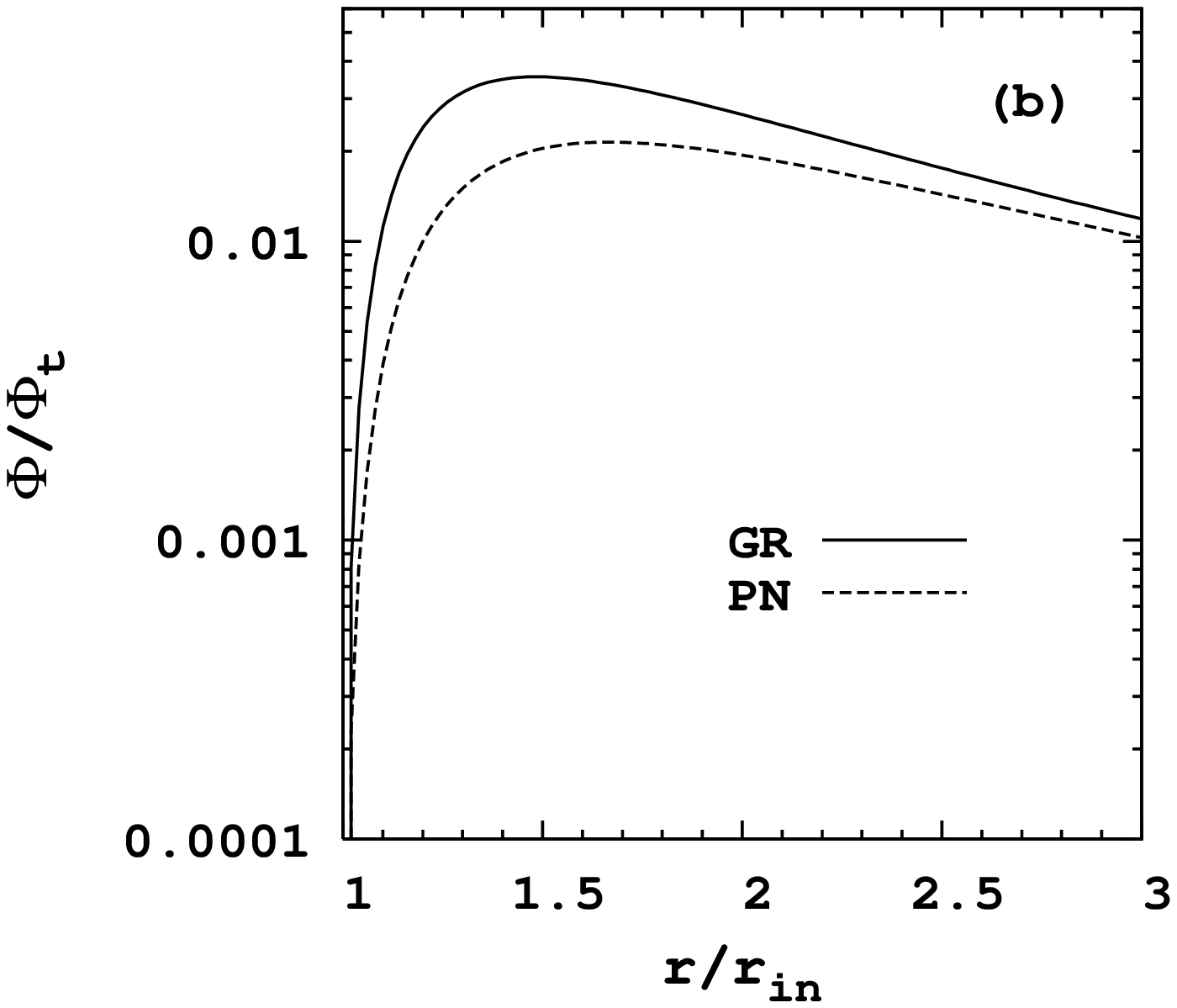} \caption{Radiation flux
emerging from the inner disk. In both the panels (a) and (b), the
solid curve represents the general relativistic (GR) estimation for
the disk flux according to Page \& Thorne (1974). The dashed curve
in both the panels (a) and (b) estimates the disk flux according to
our pseudo-Newtonian (PN) analysis. The panel (a) shows the flux
from the accretion disk around a non-rotating black hole $(a=0)$.
The disk flux for a rotating black hole with spin parameter $a=0.9$
is displayed in panel (b). \label{fig2}}
\end{figure}


\begin{figure}[htbp]
\epsscale{1.0} \plottwo{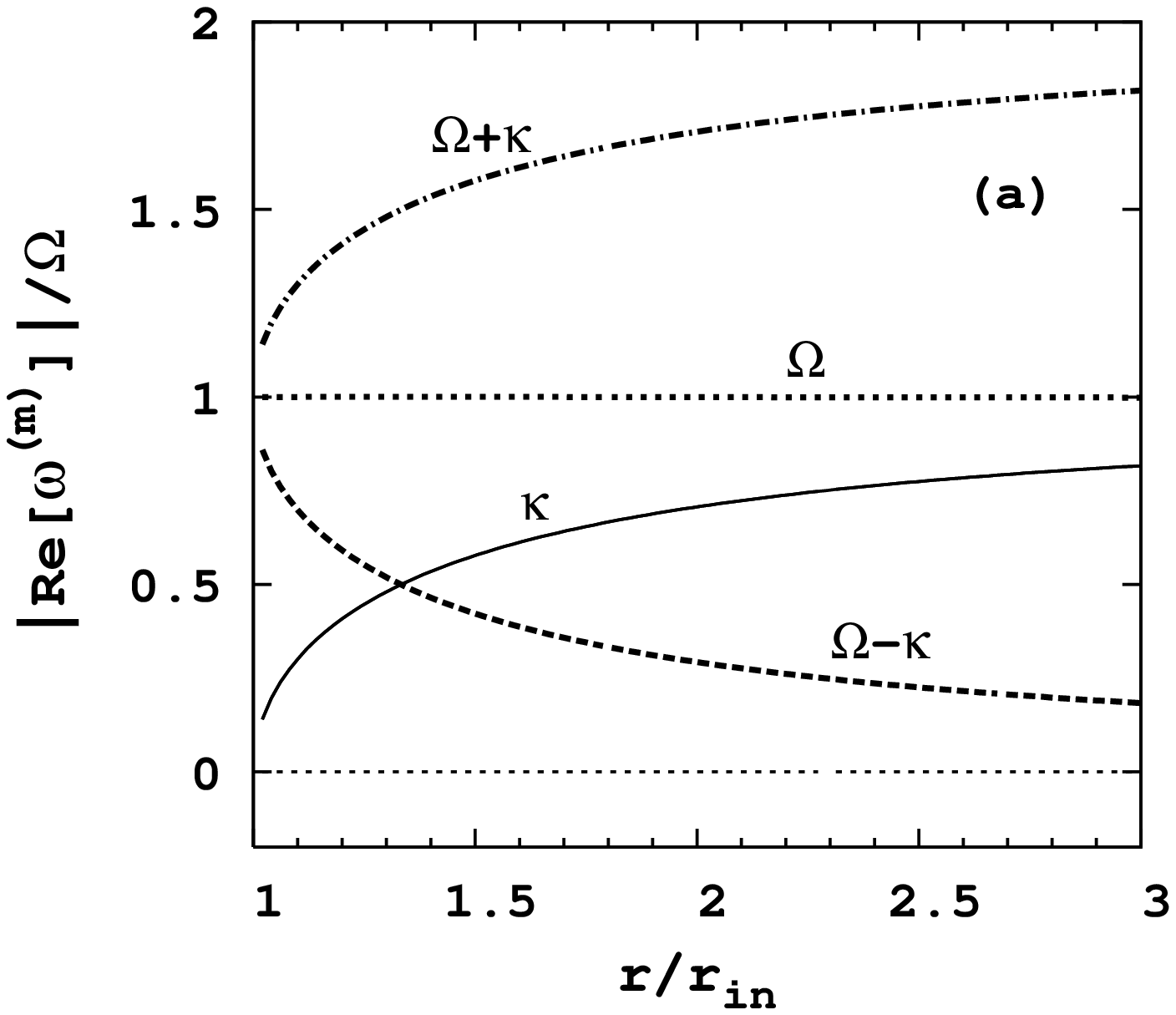}{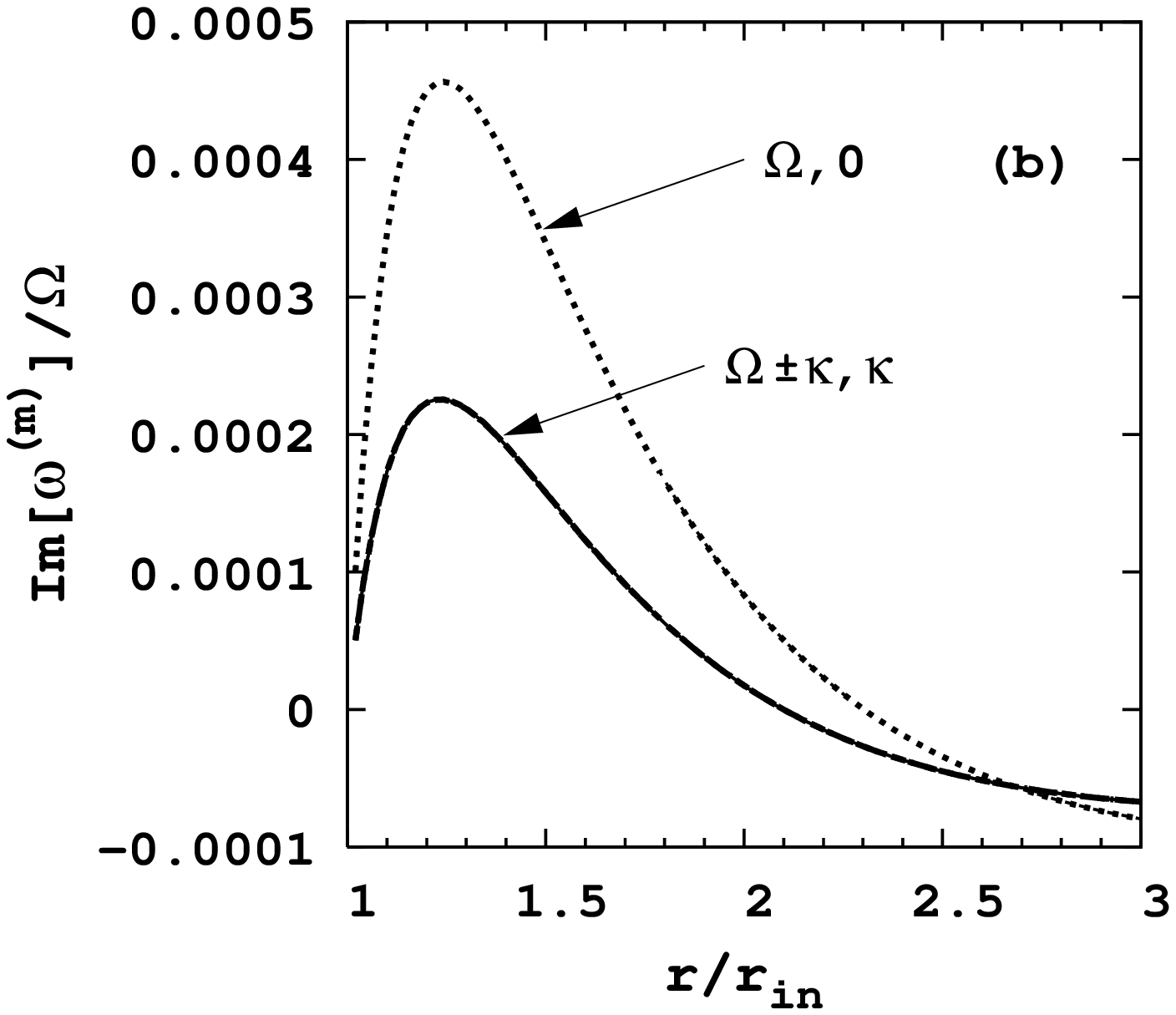} \caption{Real and
imaginary parts of the complex frequencies for axisymmetric ($m=0$)
and non-axisymmetric ($m=1$) modes in the innermost region of the
accretion disk around a non-rotating black hole $(a=0)$. The
viscosity parameter and the accretion rate are $\alpha =0.1$ and
$\dot{m}=0.1$, respectively. The mode frequencies in panel (a) and
the growth rates of the modes in panel (b) are labeled with the
corresponding test-particle frequency branches. \label{fig3}}
\end{figure}


\begin{figure}[htbp]
\epsscale{1.0} \plottwo{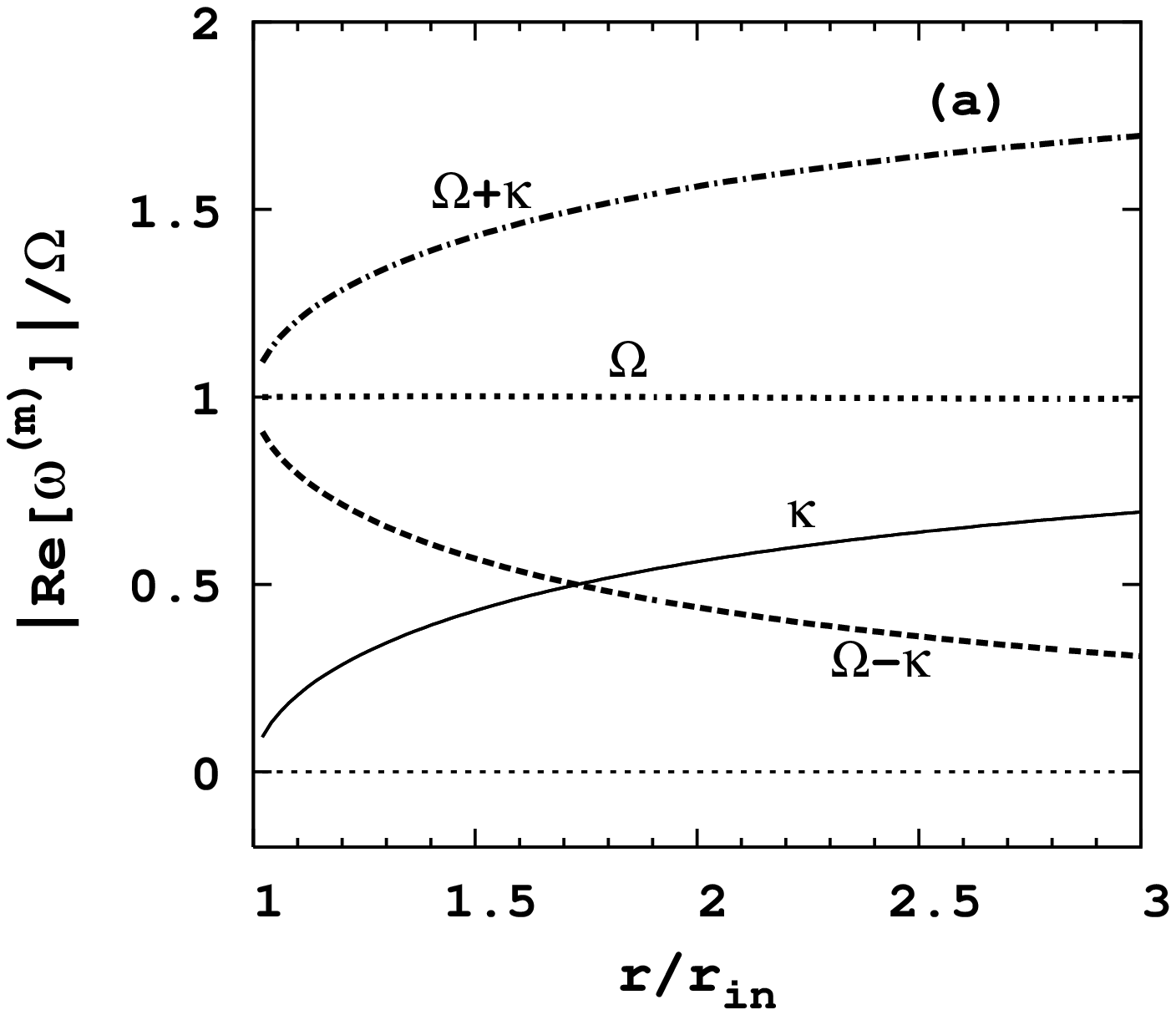}{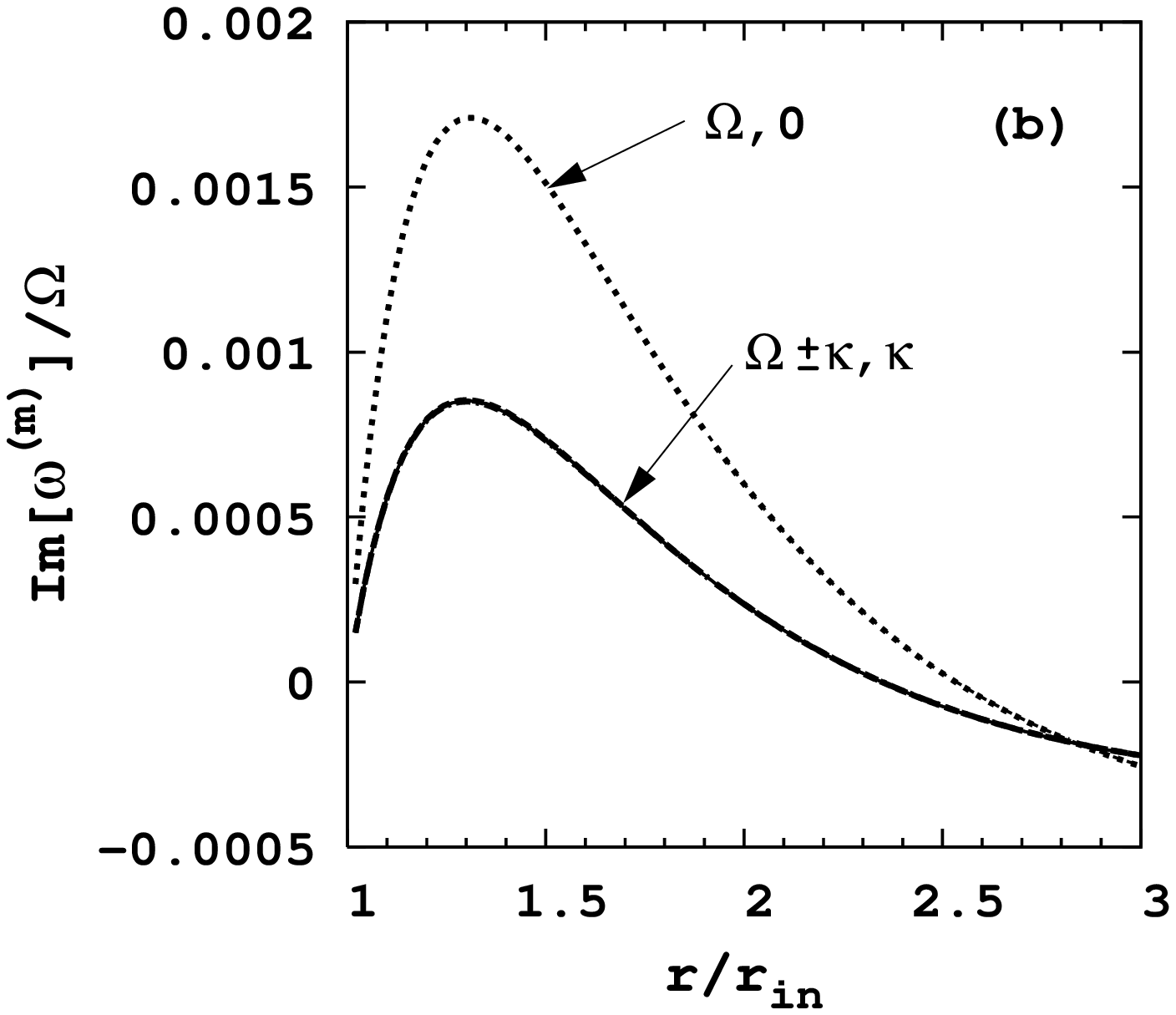} \caption{Real and
imaginary parts of the complex frequencies for axisymmetric ($m=0$)
and non-axisymmetric ($m=1$) modes in the innermost region of the
accretion disk around a rotating black hole with spin parameter
$a=0.9$. The viscosity parameter and the accretion rate are $\alpha
=0.1$ and $\dot{m}=0.1$, respectively. The mode frequencies in panel
(a) and the growth rates of the modes in panel (b) are labeled with
the corresponding test-particle frequency branches. \label{fig4}}
\end{figure}


\begin{figure}[htbp]
\epsscale{1.0} \plottwo{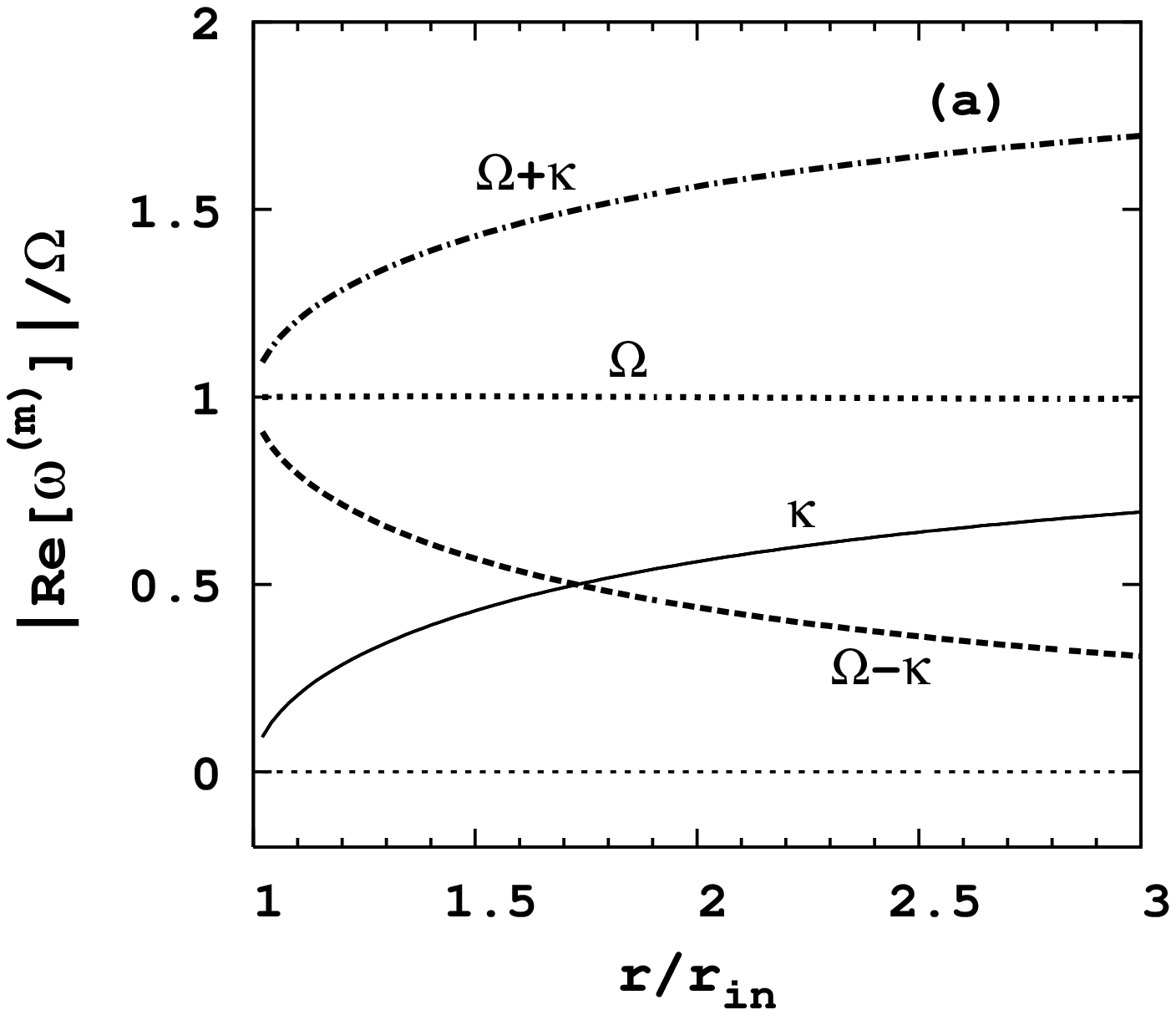}{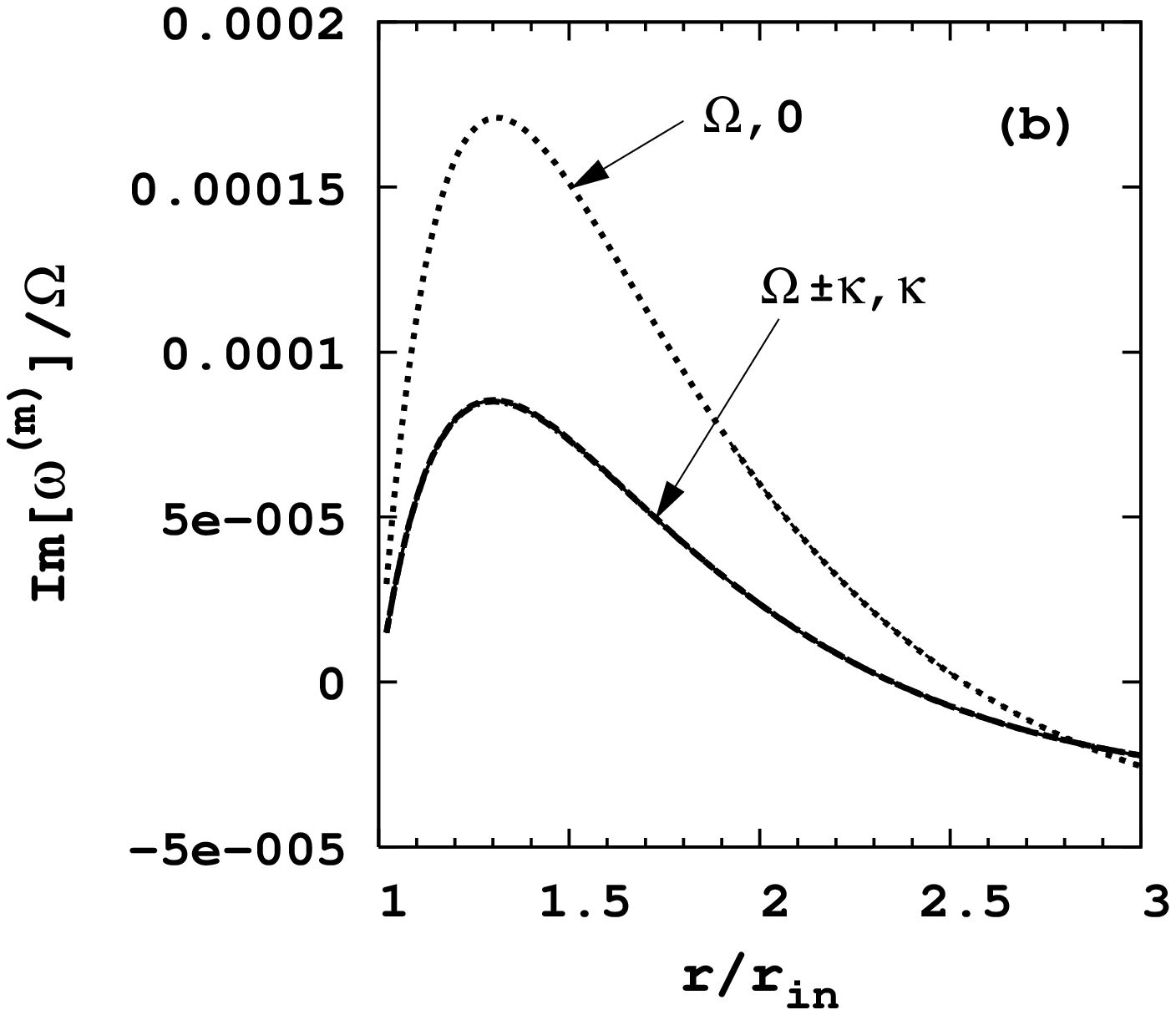} \caption{Real and
imaginary parts of the complex frequencies for axisymmetric ($m=0$)
and non-axisymmetric ($m=1$) modes in the innermost region of the
accretion disk around a rotating black hole with spin parameter
$a=0.9$. The viscosity parameter and the accretion rate are $\alpha
=0.01$ and $\dot{m}=0.1$, respectively. The mode frequencies in
panel (a) and the growth rates of the modes in panel (b) are labeled
with the corresponding test-particle frequency branches.
\label{fig5}}
\end{figure}


\begin{figure}[htbp]
\epsscale{1.0} \plottwo{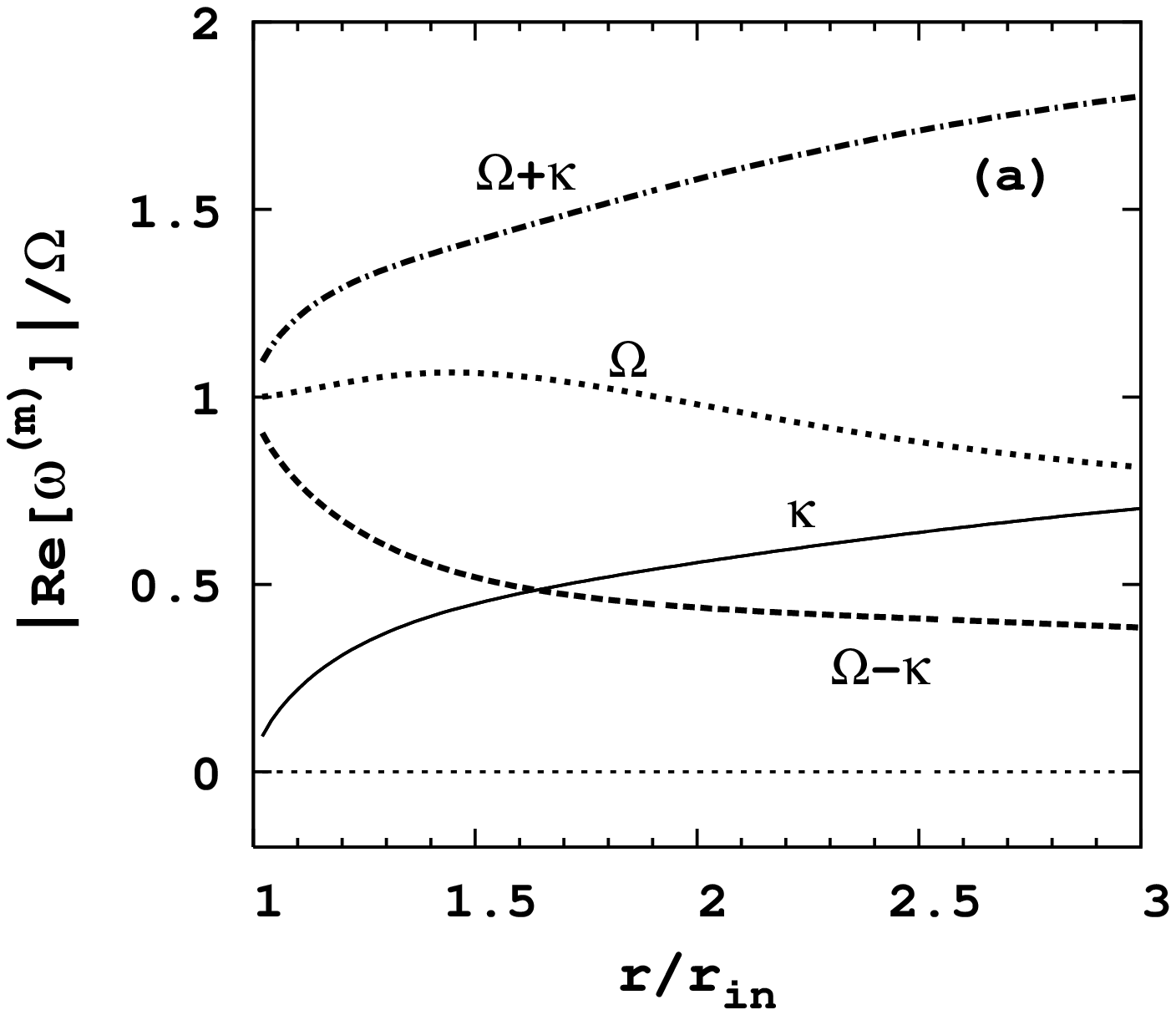}{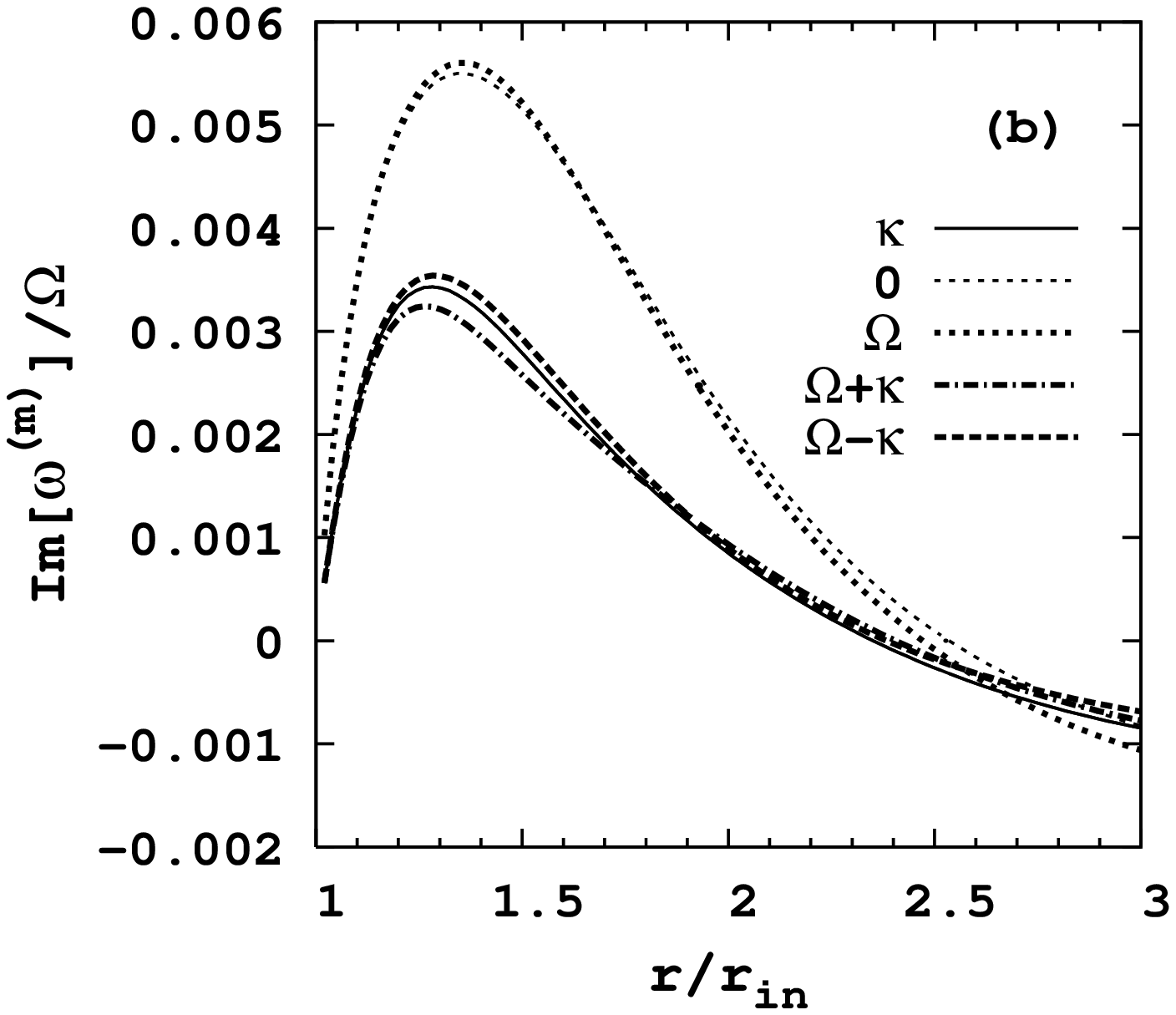} \caption{Real and
imaginary parts of the complex frequencies for axisymmetric ($m=0$)
and non-axisymmetric ($m=1$) modes in the innermost region of the
accretion disk around a rotating black hole with spin parameter
$a=0.9$. The viscosity parameter and the accretion rate are $\alpha
=0.01$ and $\dot{m}=0.6$, respectively. The mode frequencies in
panel (a) and the growth rates of the modes in panel (b) are labeled
with the corresponding test-particle frequency branches.
\label{fig6}}
\end{figure}


\begin{figure}[htbp]
\epsscale{0.5} \plotone{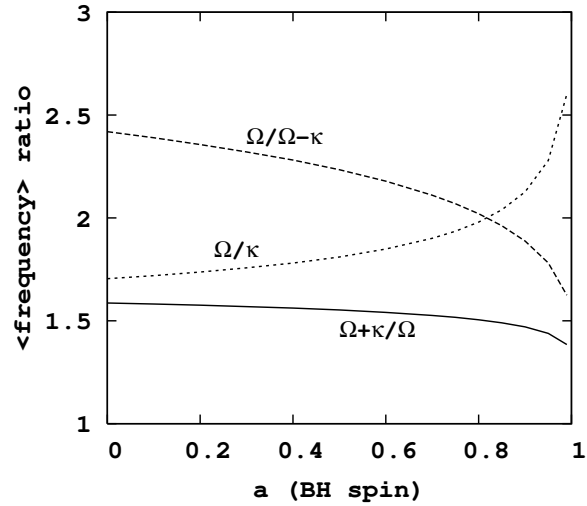} \caption{Ratios of the flux weighted
averages of different frequency bands of the modes growing in the
inner disk for all possible values of the spin parameter $a$ between
$0$ and $1$. The frequency ratios as functions of the spin parameter
are obtained for the typical values, $\alpha =0.1$ and $\dot{m}=0.1
$. \label{fig7}}
\end{figure}


\begin{figure}[htbp]
\epsscale{0.8} \plotone{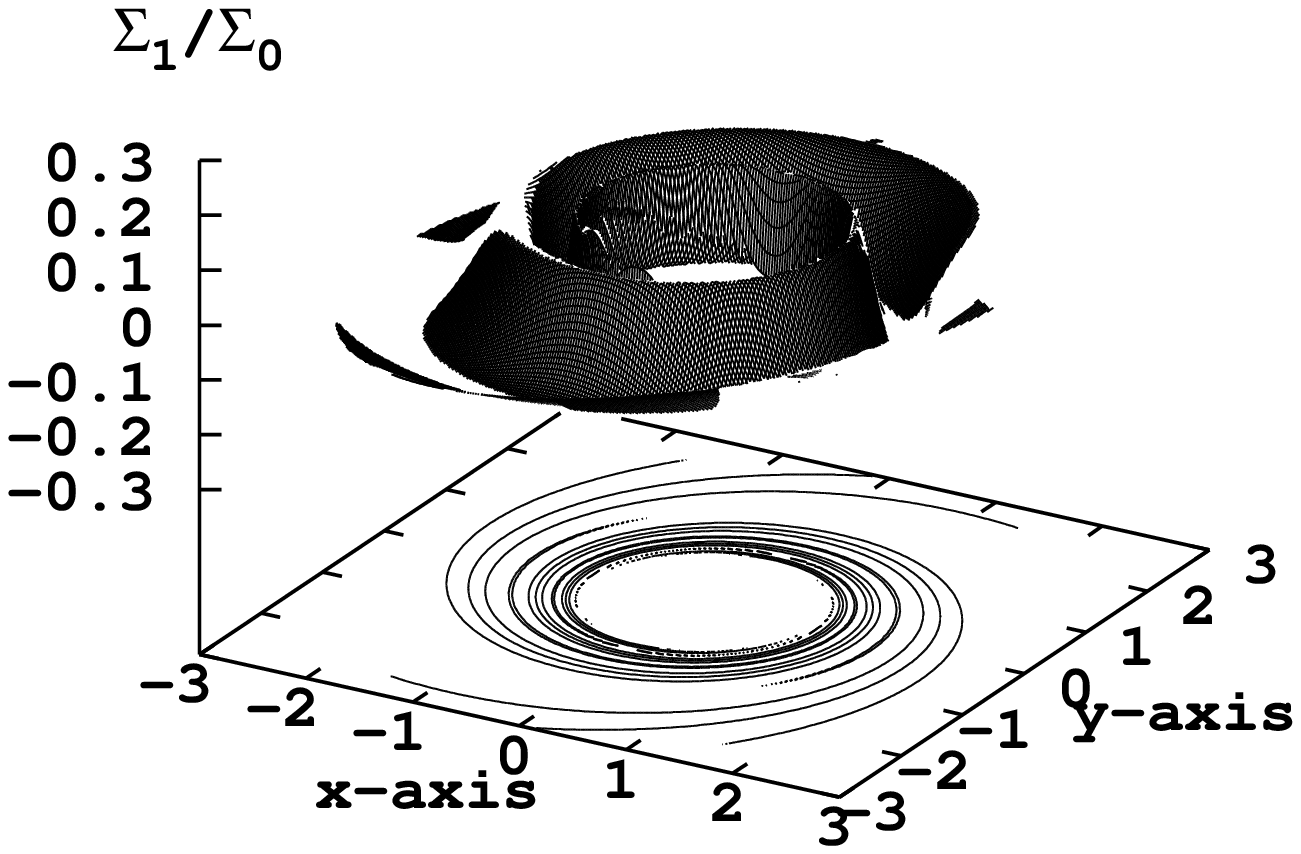} \caption{Profile of surface density
perturbation $\Sigma _{1}$ with respect to background surface
density $\Sigma _{0}$ for nonaxisymmetric mode with frequency
$\Omega $ in the innermost region ($r_{\mathrm{in}}\leq r\leq
3r_{\mathrm{in}}$) of a disk around a rotating black hole with spin
parameter $a=0.9$ for the typical values, $\alpha =0.1$ and
$\dot{m}=0.1$. The profile is obtained at the time $t=2P_{%
\mathrm{in}}$, where $P_{\mathrm{in}}=2\pi /\Omega
(r_{\mathrm{in}})$ is the rotation period at the innermost disk
radius. \label{fig8}}
\end{figure}


\begin{figure}[htbp]
\epsscale{0.8} \plotone{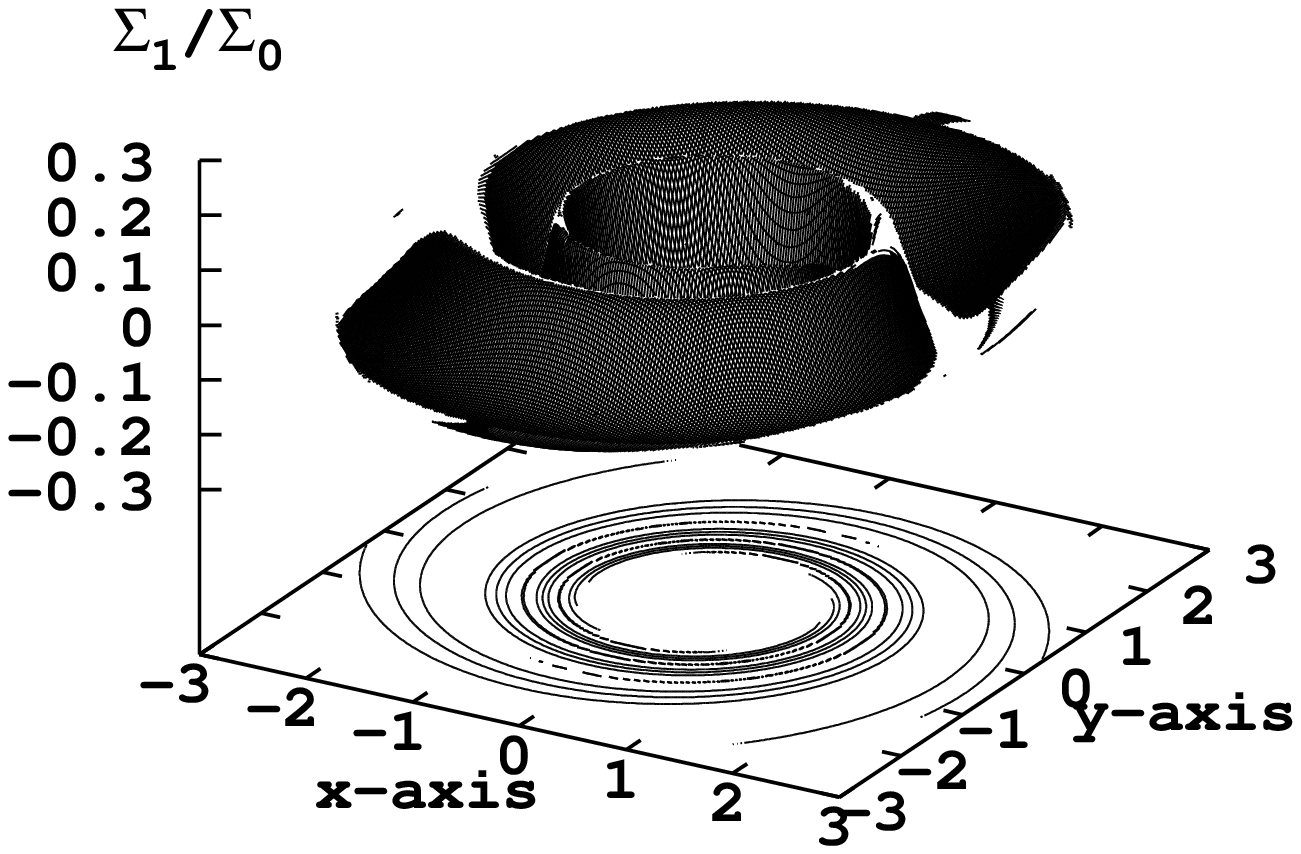} \caption{Profile of surface density
perturbation $\Sigma _{1}$ with respect to background surface
density $\Sigma _{0}$ for nonaxisymmetric mode with frequency
$\Omega +\kappa $ in the innermost region ($r_{\mathrm{in}}\leq
r\leq 3r_{\mathrm{in}}$) of a disk around a rotating black hole with
spin parameter $a=0.9$ for the typical values, $\alpha =0.1$ and
$\dot{m}=0.1$. The profile is obtained at the time $t=2P_{%
\mathrm{in}}$, where $P_{\mathrm{in}}=2\pi /\Omega
(r_{\mathrm{in}})$ is the rotation period at the innermost disk
radius. \label{fig9}}
\end{figure}


\begin{figure}[htbp]
\epsscale{0.8} \plotone{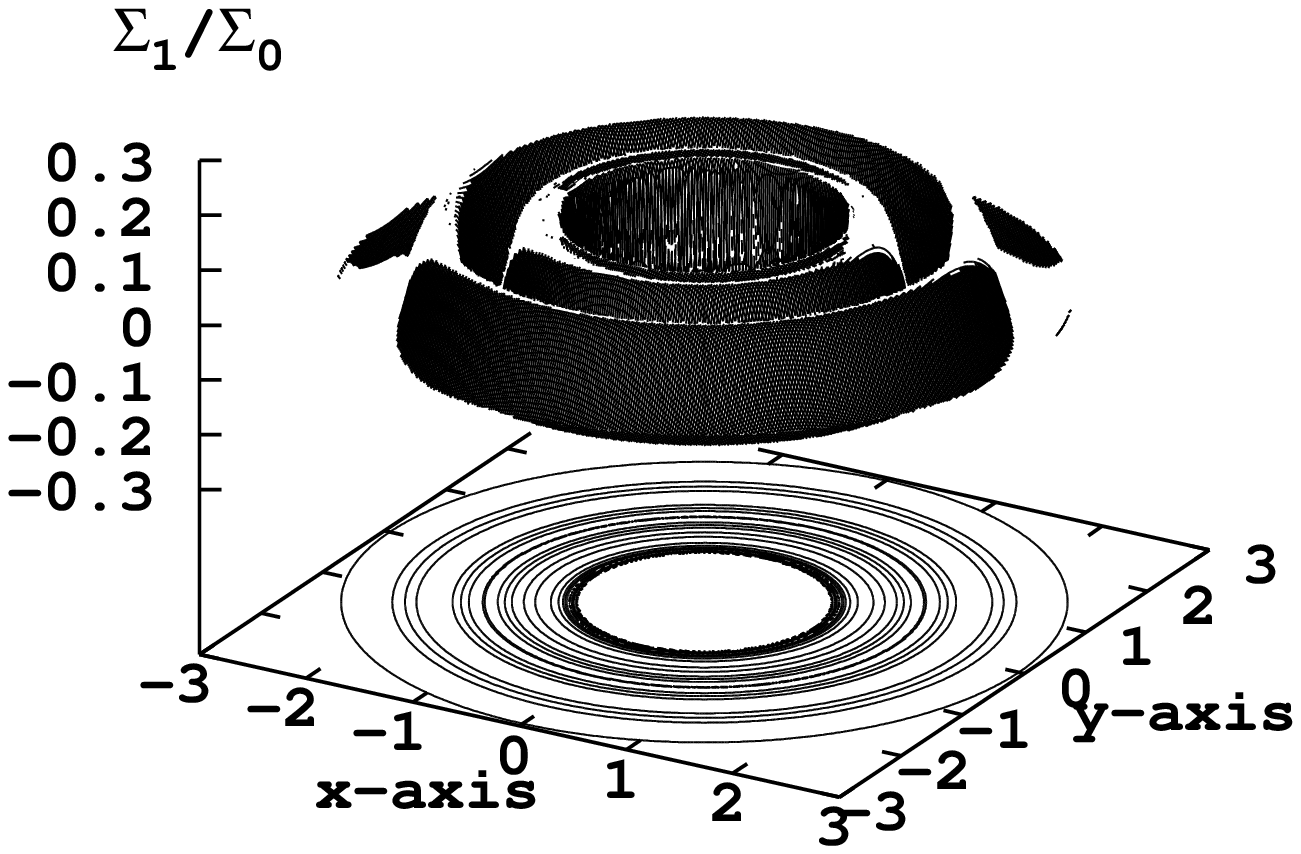} \caption{Profile of surface density
perturbation $\Sigma _{1}$ with respect to background surface
density $\Sigma _{0}$ for axisymmetric mode with frequency $\kappa $
in the innermost region ($r_{\mathrm{in}}\leq r\leq
3r_{\mathrm{in}}$) of a disk around a rotating black hole with spin
parameter $a=0.9$ for the typical values, $\alpha =0.1$ and
$\dot{m}=0.1$. The profile is obtained at the time $t=15P_{%
\mathrm{in}}$, where $P_{\mathrm{in}}=2\pi /\Omega
(r_{\mathrm{in}})$ is the rotation period at the innermost disk
radius. \label{fig10}}
\end{figure}


\begin{figure}[htbp]
\epsscale{0.8} \plotone{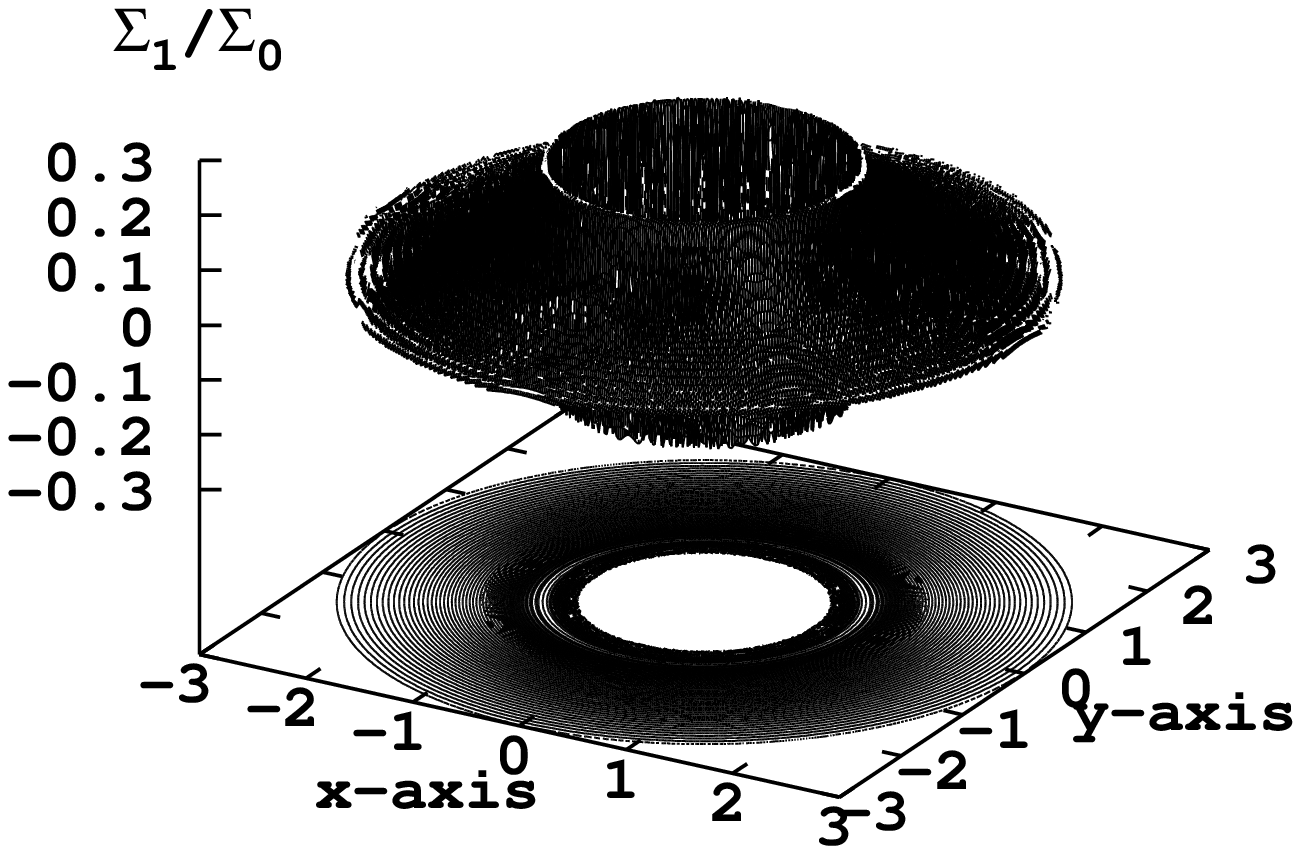} \caption{Profile of surface density
perturbation $\Sigma _{1}$ with respect to background surface
density $\Sigma _{0}$ for axisymmetric mode with frequency $\kappa $
in the innermost region ($r_{\mathrm{in}}\leq r\leq
3r_{\mathrm{in}}$) of a disk around a rotating black hole with spin
parameter $a=0.9$ for the typical values, $\alpha =0.1$ and
$\dot{m}=0.1$. The profile is obtained at the time $t=200P_{%
\mathrm{in}}$, where $P_{\mathrm{in}}=2\pi /\Omega
(r_{\mathrm{in}})$ is the rotation period at the innermost disk
radius. \label{fig11}}
\end{figure}

\end{document}